\documentclass[11pt]{article}
\pdfoutput=1 
\usepackage{mathtools}
\usepackage{booktabs}
\usepackage{amsmath,amssymb,amsbsy,amstext, amsthm, simplewick, amsfonts,braket}
\usepackage{graphicx}
\usepackage[small]{caption}
\usepackage{siunitx}
\usepackage{subcaption}
\usepackage{upgreek}
\usepackage{framed}
\usepackage{wrapfig}
\usepackage{multirow}
\usepackage{bbm}
\usepackage[numbers,sort&compress]{natbib}
\usepackage[svgnames,dvipsnames,x11names]{xcolor}
\usepackage[utf8x]{inputenc}
\usepackage{selinput}
\usepackage{bm}
\usepackage{float}
\usepackage{dsfont}
\usepackage[margin=1cm]{caption}
\usepackage{subcaption}
\usepackage{sidecap}
\usepackage{longtable}
\usepackage{anyfontsize}

\usepackage{epstopdf}
\usepackage{cancel}
\usepackage{tcolorbox}
\usepackage{latexsym,amsmath,amssymb,epsfig}
\usepackage{braket}
\usepackage{tensor}
\usepackage{tocloft}
\usepackage[permil]{overpic}

\usepackage{tcolorbox}
\definecolor{greyish2}{rgb}{.96,.96,.96}

\def\xyma{\xymatrix@M.7em}
\def\xymas{\xymatrix@M.1em}

\newcommand{\Comment}[1]{{}}
\definecolor{darkblue}{rgb}{0.15,0.35,0.55}
\definecolor{reddish}{rgb}{0.65, 0.2, 0.2}
\definecolor{darkgreen}{RGB}{50,150,0}
\definecolor{greyish2}{rgb}{.96,.96,.96}
\usepackage[linktocpage=true]{hyperref}
\hypersetup{
colorlinks=true,
citecolor=darkblue,
linkcolor=reddish,
urlcolor=darkblue,
pdfauthor={},
pdftitle={},
pdfsubject={}
}

\DeclareFontFamily{OT1}{rsfs10}{}
\DeclareFontShape{OT1}{rsfs10}{m}{n}{ <-> rsfs10 }{}
\DeclareMathAlphabet{\mathscript}{OT1}{rsfs10}{m}{n}

\newcommand\blfootnote[1]{%
  \begingroup
  \renewcommand\thefootnote{}\footnote{#1}%
  \addtocounter{footnote}{-1}%
  \endgroup
}

\def\gsim{ \lower .75ex \hbox{$\sim$} \llap{\raise .27ex \hbox{$>$}} }
\def\lsim{ \lower .75ex \hbox{$\sim$} \llap{\raise .27ex \hbox{$<$}} }
\def\be{\begin{equation}}
\def\ee{\end{equation}}
\def\bea{\begin{eqnarray}}
\def\eea{\end{eqnarray}}

\def\K{K}
\newcommand{\baaa}{\begin{eqnarray}}
\newcommand{\eaaa}{\end{eqnarray}}

\newcommand{\T}{\mathcal{T}}

\DeclareMathOperator{\E}{e}

\usepackage[a4paper,margin=1in]{geometry}

\usepackage{tikz}
\usetikzlibrary{decorations}
\pgfdeclaredecoration{complete sines}{initial}
{
    \state{initial}[
        width=+0pt,
        next state=upsine,
        persistent precomputation={\pgfmathsetmacro\matchinglength{
            \pgfdecoratedinputsegmentlength / int(\pgfdecoratedinputsegmentlength/\pgfdecorationsegmentlength)}
            \setlength{\pgfdecorationsegmentlength}{\matchinglength pt}
        }] {}
    \state{upsine}[width=\pgfdecorationsegmentlength,next state=downsine]{
        \pgfpathsine{\pgfpoint{0.25\pgfdecorationsegmentlength}{0.5\pgfdecorationsegmentamplitude}}
        \pgfpathcosine{\pgfpoint{0.25\pgfdecorationsegmentlength}{-0.5\pgfdecorationsegmentamplitude}}
    }
    \state{downsine}[width=\pgfdecorationsegmentlength,next state=upsine]{
        \pgfpathsine{\pgfpoint{0.25\pgfdecorationsegmentlength}{-0.5\pgfdecorationsegmentamplitude}}
        \pgfpathcosine{\pgfpoint{0.25\pgfdecorationsegmentlength}{0.5\pgfdecorationsegmentamplitude}}
}
    \state{final}{}
}

\definecolor{greyish}{rgb}{.90,.90,.90}
\definecolor{greyish2}{rgb}{.96,.96,.96}
\usepackage{xcolor,colortbl}
\usepackage{tcolorbox}

\usepackage[all]{xy}

\def \v {\nu}
\def \lll {\langle\!\langle}
\def \rr{\rangle\!\rangle}

\newcommand{\p}{\partial}
\renewcommand{\O}{\mathcal{O}}
\renewcommand{\E}{\mathcal{E}}

\DeclareSymbolFont{matha}{OML}{txmi}{m}{it}
\DeclareMathSymbol{v}{\mathord}{matha}{118}

\newcommand{\sdot}{\!\cdot\!}
\newcommand{\del}{\nabla}
\renewcommand{\O}{{\mathcal O}}
\newcommand{\sep}{{\rm sep}}

\newcommand{\tT}{\widetilde{T}}
\newcommand{\tTheta}{\widetilde{\Theta}}

\newcommand{\tO}{\widetilde{O}}
\newcommand{\R}{{\cal R}}

\newcommand{\bz}{\bar{z}}


\numberwithin{equation}{section}
\begin{document}
%
\renewcommand{\thefootnote}{\fnsymbol{footnote}}
\vspace{0truecm}
\thispagestyle{empty}

\begin{center}
{
\bf\LARGE
Averaged Null Energy and the \\ 
\bigskip
Renormalization Group 
}
\end{center}

\vspace{.15truecm}

\begin{center}
{\fontsize{12.7}{18}\selectfont
Thomas Hartman\blfootnote{\texttt{\href{mailto:hartman@cornell.edu}{hartman@cornell.edu}}}
and Gr\'egoire Mathys\blfootnote{\texttt{\href{mailto:gregoire.mathys@cornell.edu}{gregoire.mathys@cornell.edu}}} 
}

%

\vspace{0.7cm}

{\it Department of Physics} \\
{\it Cornell University, Ithaca, NY 14850, USA}

\end{center}

\vspace{0.7cm}

\begin{abstract}
\noindent
We establish a connection between the averaged null energy condition (ANEC) and the monotonicity of the renormalization group, by studying the light-ray operator $\int du T_{uu}$ in quantum field theories that flow between two conformal fixed points. In four dimensions, we derive an exact sum rule relating this operator to the Euler coefficient in the trace anomaly, and show that the ANEC implies the $a$-theorem. The argument is based on matching anomalies in the stress tensor 3-point function, and relies on special properties of contact terms involving light-ray operators. We also illustrate the sum rule for the example of a free massive scalar field. 
Averaged null energy appears in a variety of other applications to quantum field theory, including causality constraints, Lorentzian inversion, and quantum information. The quantum information perspective provides a new derivation of the $a$-theorem from the monotonicity of relative entropy. The equation relating our sum rule to the dilaton scattering amplitude in the forward limit suggests an inversion formula for non-conformal theories.

\end{abstract}

\newpage

\setcounter{page}{2}
\setcounter{tocdepth}{2}
\tableofcontents
\newpage
\renewcommand*{\thefootnote}{\arabic{footnote}}
\setcounter{footnote}{0}




\section{Introduction}

The averaged null energy (ANE) is the nonlocal operator 
\begin{align}
\E_u = \int du T_{uu} \ ,
\end{align}
where $u$ is a null coordinate. 
In the last few years it has become clear that this and similar operators, known generally as light-ray operators, play an interesting role in quantum field theory and quantum gravity. They measure the flow of energy and other quantities, thus providing infrared-safe observables that are complementary to the $S$-matrix \cite{Hofman:2008ar,Dixon:2019uzg,Kologlu:2019mfz,Lee:2022ige}. In conformal field theory, light-ray operators control the behavior of certain correlation functions, because they appear in the lightcone and Regge OPEs and are often the leading contributions to the Lorentzian inversion formula \cite{Hartman:2016lgu,Caron-Huot:2017vep,Simmons-Duffin:2017nub}. The ANE also appears in deformations of the modular Hamiltonian \cite{Faulkner:2016mzt}, determines the onset of quantum chaos \cite{Hartman:2015lfa,Hartman:2016lgu,Kundu:2021qcx}, and is closely tied to soft theorems and asymptotic symmetries \cite{Casini:2017roe,Cordova:2018ygx,Hu:2022txx}. In holographic theories the ANE is dual to a gravitational shockwave and tied to the emergence of causality \cite{Hofman:2008ar,Hofman:2009ug,Kelly:2014mra}. 
In quantum field theory in Minkowski spacetime, the ANE satisfies the positivity condition $\langle\Psi| \E_u|\Psi \rangle \geq 0$ in any state.  
This is the averaged null energy condition, or ANEC. The ANEC was derived in various free theories long ago \cite{Klinkhammer:1991ki,Wald:1991xn,Folacci:1992xg,Ford:1995gb} and more recently extended to interacting theories using the monotonicity of relative entropy \cite{Faulkner:2016mzt}, and separately, conformal bootstrap methods together with reflection positivity \cite{Hartman:2016lgu}. In a theory coupled to gravity, a suitable generalization of the ANEC underlies a variety of classic theorems on causality, singularities, positive energy, and wormholes \cite{Borde_1987,Gao:2000ga,Graham_2007}. In quantum field theory without gravity, the ANEC constrains the coupling constants and anomaly coefficients of CFTs \cite{Hofman:2008ar,Hofman:2009ug} (see also \cite{Hartman:2016lgu,Hartman:2015lfa,Hartman:2016dxc,Hofman:2016awc,Cordova:2017zej,Bautista:2019qxj}); its higher spin relatives constrain the 3D critical Ising model \cite{Hartman:2016lgu}, a prediction that was subsequently verified by numerical bootstrap \cite{Simmons-Duffin:2016wlq}; and through holographic duality, the ANEC was an early hint for how to derive Einstein gravity from CFT \cite{Afkhami-Jeddi:2016ntf,Meltzer:2017rtf,Belin:2019mnx,Kologlu:2019bco,Belin:2020lsr}. These techniques have now been subsumed by the Lorentzian inversion formula in the program to derive strict inequalities for CFT correlators and scattering amplitudes \cite{Caron-Huot:2020adz,Caron-Huot:2021enk,Caron-Huot:2021rmr,Caron-Huot:2022ugt}. 

In this paper, we will study the interplay between the ANE and the renormalization group (RG). We consider quantum field theories in Minkowski spacetime in four dimensions. In CFTs, we show that correlators of $\E_u$ have contact terms dictated by the conformal anomaly. In non-conformal QFTs, we derive sum rules for the change in the Euler anomaly along an RG flow. The sum rules, together with the ANEC, provide a new derivation of the $a$-theorem. Similar methods can be used to derive the Zamolodchikov $c$-theorem from the ANEC in two dimensions; the application to two dimensions will be reported in a separate paper \cite{comingSoon} together with other constraints from the two-dimensional ANEC.

These two theorems establish that the renormalization group is monotonic in $d=2$ and $d=4$. The $c$-theorem was derived by Zamolodchikov using the stress-tensor two-point function $\langle T_{\alpha\beta} T_{\rho\sigma} \rangle$ at separated points \cite{Zamolodchikov:1986gt}. The $a$-theorem, conjectured by Cardy \cite{Cardy:1988cwa}, was derived by Komargodski and Schwimmer \cite{Komargodski:2011vj} using background field techniques and anomaly matching applied to the contact terms in the $\langle \Theta\Theta\Theta\Theta\rangle$ correlation function, where $\Theta = T\indices{_\alpha^\alpha}$ is the trace of the stress tensor. 
Both of these theorems have also been derived from quantum information theory \cite{Casini:2004bw,Casini:2006es,Casini:2012ei,Casini:2017vbe,Casini:2023kyj}.  

Nonetheless, the physical interpretation of these results, known collectively as $C$-theorems, remains quite mysterious. It is often said that the $C$-theorems come from the reduction in degrees of freedom as massive states are integrated out along an RG flow. But this is unsatisfying: Why should the Euler anomaly measure degrees of freedom in even dimensions? And is there a $C$-theorem in five or more dimensions? The entropic proof applies in three dimensions \cite{Casini:2012ei}, but has not been extended to five or more, and it relies on subleading terms in the entanglement entropy so there is still no clear connection to counting degrees of freedom. The situation in holographic theories is better, because there is a universal, dimension-independent derivation of the $C$-theorems in the bulk based on the null energy condition for bulk matter \cite{Myers:2010xs}. But the dual of this argument, phrased in the language of the boundary QFT, is unknown --- the dual of the bulk null energy condition is not the boundary null energy condition \cite{Hartman:2022njz}. It thus remains a fascinating problem to understand the $C$-theorems more deeply.

Our approach is based on the 3-point correlation function $\langle \Theta \E_u \Theta\rangle$ in both two and four dimensions. 
In quantum field theories that flow between two conformal fixed points, we derive the following sum rules for the change in the Euler anomaly from the UV to the IR. In four dimensions, in the coordinates  $ds^2 = -du dv + d\vec{x}^2$, 
\begin{align}\label{introsum4}
a_{UV} - a_{IR}  =- \frac{1}{32} \int_{v_1<0}d^4x_1\int_{v_2<0} d^4 x_2\, (u_1-u_2)^2 \vec{x}_1 \cdot \vec{x}_2 \langle \Theta(x_1) T_{uu}(0) \Theta(x_2)\rangle \ .
\end{align}
The application to two dimensions will be discussed in a separate paper \cite{comingSoon}. The analogous sum rule is 
\begin{align}\label{introsum2}
c_{UV} - c_{IR}  = -6\pi  \int_{v_1<0}d^2x_1\int_{v_2<0} d^2 x_2\, (u_1-u_2)^2 \langle \Theta(x_1)T_{uu}(0) \Theta(x_2)\rangle\, .
\end{align}
The 3-point functions in the integrands are Wightman functions, with no contact terms. 
These integrals can be rewritten as expectation values of the averaged null energy, $\langle \Psi | \int du T_{uu} | \Psi\rangle$.  Therefore, the ANEC implies that both $\Delta c$ and $\Delta a$ are non-negative. The state $|\Psi\rangle$ is defined by an insertion of the trace $\Theta$ smeared against a particular kernel (see \eqref{positivesumrule4d}). It is similar to the `conformal collider' state studied by Hofman and Maldacena in CFT \cite{Hofman:2008ar}, except that in our case, the wavepacket is inserted close to the ANE light ray rather than at large separation.

The sum rules can also be written in terms of time-ordered or retarded correlation functions, and the latter version provides a physical interpretation: They measure the response of $\langle \E_u\rangle$ to a Weyl deformation of the Hamiltonian.

The logic behind \eqref{introsum4}-\eqref{introsum2} is simple once we have developed some new techniques for the analysis of light-ray operators in QFT. In particular we study the role of anomalies and contact terms in correlation functions involving a light-ray operator, and constrain the types of contact terms that can be generated along an RG flow. 
Other RG sum rules for stress tensor correlators have been discussed in the literature \cite{Adler:1982ri,Zee:1980sj,Anselmi:2001yp}. An important difference in \eqref{introsum4}-\eqref{introsum2} is that the integrals have no contact terms and can be expressed in a way that is manifestly positive. 
The holographic RG sum rules derived in \cite{Baumann:2019ghk} are also related to the boundary ANEC, and they are based on the same intuition that $C$-theorems should be related to the spreading of quantum information, but they have a more limited regime of validity.

At present we have not found a way to extend our methods to odd dimensions, or to $d=6$, where there is a supersymmetric $a$-theorem \cite{Heckman:2015axa,Cordova:2015fha,Stergiou:2016uqq,Cordova:2020tij,Heckman:2021nwg} but the general case remains open.

The averaged null energy operator plays an important role in two other recent developments in quantum field theory: quantum information of deformed half-spaces \cite{Faulkner:2016mzt,Casini:2017roe}, and the Lorentzian inversion formula \cite{Caron-Huot:2017vep}. Thus our results connect these developments to the RG. In the discussion section we point out the connections, leaving a detailed exploration to future work. It is unclear exactly how our results are related to the derivation of the $a$-theorem by Komargodski and Schwimmer, but we suggest that the relation should be viewed as a generalization of the Lorentzian inversion formula to non-conformal theories. We also refer the reader to the discussion section for a brief technical summary that outlines the derivation of the sum rule.

\subsubsection*{Outline}

We setup our conventions and discuss some relevant properties of the ANE in section \ref{sec:prelim}.  We then turn to a general discussion of contact terms in Lorentzian signature,  and at a conformal fixed point in $d=4$  dimensions, derive the contact terms in the retarded correlator $\langle \R[ \E_u ; \Theta\Theta]\rangle$ and the time-ordered correlator $\langle {\cal T}[\E_u \Theta\Theta]\rangle$ (section \ref{sec:anomaly}). In section \ref{sec:atheorem} we consider quantum field theories that flow between fixed points, apply the CFT results to the infrared fixed point, and derive the sum rules and the $a$-theorem. Finally we work out the example of a free massive scalar in section \ref{sec:MassiveScalar}. In the discussion (section \ref{s:discussion}), we summarize the strategy used to the derive the sum rule and discuss its interpretation from the perspectives of quantum information and, more speculatively, Lorentzian inversion.

\section{Preliminaries\label{sec:prelim}}

In this section, we present the relevant material for the rest of this work, including our conventions for the stress tensor and correlation functions. We then discuss properties of the ANE operator and its behavior inside correlation functions.

\subsection{Conventions}

In Euclidean spacetime we use coordinates
\be 
d s^2 = d \tau^2 + d y^2 + d \vec{x}^2 \, ,
\ee
with $\vec{x}\in \mathbb{R}^{d-2}$. In Minkowski spacetime, we use 
\be 
d s^2 = -du dv + d \vec{x}^2 = -d t^2 + d y^2 + d \vec{x}^2\, ,
\ee
where 
\be 
u = t-y = -(y+i\tau)\, ,\qquad \qquad \v = t + y =y-i\tau \, .
\ee
%
We define the stress tensor with the convention
\begin{align}
\label{eq:DefT}
T_{\mu\nu} \equiv \frac{-2}{\sqrt{-g}}\frac{\delta S}{\delta g^{\mu\nu}} \ , 
\qquad
\langle T_{\mu\nu}\rangle = \frac{2i}{\sqrt{-g}} \frac{\delta}{\delta g^{\mu\nu} }\log Z \ .
\end{align}
In Euclidean signature this convention corresponds to
\begin{align}
T_{\mu\nu} \equiv \frac{2}{\sqrt{g}}\frac{\delta S_E}{\delta g^{\mu\nu}}\, , \qquad\qquad 
\langle T_{\mu\nu}\rangle = -\frac{2}{\sqrt{g}} \frac{\delta}{\delta g^{\mu\nu}} \log Z\, ,
\end{align}
with $S_E$ the Euclidean action.
%
%
%
The trace of the stress tensor is denoted 
\be 
\Theta  = T_\mu^\mu = \frac{2}{\sqrt{-g}}\frac{\delta S}{\delta \omega} = -\frac{2}{\sqrt{g}}\frac{\delta S_E}{\delta \omega} \, ,
\ee
with the Weyl variation 
\begin{align}
\delta g_{\mu\nu} = g_{\mu\nu} \delta \omega \ .
\end{align}
Dirac delta functions are defined covariantly such that $\int d^d x \sqrt{g}\delta^{(d)} (x) = 1$, and we use the following Fourier transform conventions:
 \be 
f(\K) = \int d^d x e^{i\K \cdot x} f(x)\, ,\qquad \qquad   f(x) = \int \frac{d^dp}{(2\pi)^d}e^{-i\K \cdot x}\, .
\ee
We denote Euclidean momentum by  $\K^\mu$ and Minkowski momentum by $k^\mu$. We use the customary double bracket notation to strip off momentum conserving delta function, 
%
%
\be 
\braket{\mathcal{O}(k_1)\cdots\mathcal{O}(k_n)} \equiv (2\pi)^d \delta^{(d)}(k_1+\cdots+k_n)\lll \mathcal{O}(k_1)\cdots\mathcal{O}(k_n)\rr\, .\label{eq:DBracket}
\ee

\subsection{Correlation functions\label{sec:DefCor}}

We will work exclusively with connected correlators, which are denoted  $\langle \cdots \rangle$.
In Euclidean signature, the stress tensor correlators are given by the variation
\begin{align}
\langle T_{\mu\nu}(x_1)\dots T_{\alpha\beta}(x_n)\rangle
&= \frac{(-2)^n}{\sqrt{g(x_1)}\dots\sqrt{g(x_n)}} \frac{\delta^n }{\delta g^{\mu\nu}(x_1)\dots \delta g^{\alpha\beta}(x_n)}\log Z\, .\label{eq:Tcorrelator1}
\end{align}
Correlators involving the trace $\Theta$ are defined with the traces taken inside the variation,
\begin{align}\label{ThetaDefCor}
\langle \Theta(x_1)\dots \Theta(x_n)\rangle
&= \frac{(-2)^n}{\sqrt{g(x_1)}\dots\sqrt{g(x_n)}}
g^{\mu\nu}(x_1) \frac{\delta}{\delta g^{\mu\nu}(x_1)}
\cdots
g^{\alpha\beta}(x_n) \frac{\delta}{\delta g^{\alpha\beta}(x_n)} \log Z \notag\\
&= \frac{2^n}{\sqrt{g(x_1)}\dots\sqrt{g(x_n)}} \frac{\delta^n}{\delta \omega(x_1) \dots \delta \omega(x_n)}\log Z \ .
\end{align}
To compute Euclidean correlators with both $T_{\mu\nu}$ and $\Theta$, the variation leading to $T_{\mu\nu}$ is done last, regardless of the order in which the fields are written:
\small
\begin{align}\label{mixedDefs}
\langle \Theta(x_1)T_{\mu\nu}(x_2)\rangle
&= \frac{4}{\sqrt{g(x_1)}\sqrt{g(x_2)}} \frac{\delta}{\delta g^{\mu\nu}(x_2)} g^{\alpha\beta}(x_1) \frac{\delta}{\delta g^{\alpha\beta}(x_1)} \log Z\\
\langle \Theta(x_1)\Theta(x_2)T_{\mu\nu}(x_3)\rangle
&= \frac{-8}{\sqrt{g(x_1)} \sqrt{g(x_2)} \sqrt{g(x_3)}} \frac{\delta}{\delta g^{\mu\nu}(x_3)}
g^{\alpha\beta}(x_2)\frac{\delta}{\delta g^{\alpha\beta}(x_2)} g^{\rho\sigma}(x_1) \frac{\delta}{\delta g^{\rho\sigma}(x_1)} \log Z \ . \notag
\end{align}
\normalsize
With these conventions, $g^{\mu\nu}\langle T_{\mu\nu} \cdots \rangle$ and $\langle \Theta\cdots\rangle$ differ by contact terms.  
The Ward identities, taking into account these definitions, imply 
\begin{align}\label{wardidentity1}
\del^\mu\langle T_{\mu\nu}(x)\Theta(x_1)\rangle &= \langle \Theta(x)\rangle \del_\nu \delta^{(d)}(x-x_1)\, ,
\end{align}
and
\begin{align}\label{wardidentity2}
\del^\mu\langle T_{\mu\nu}(x)\Theta(x_1)\Theta(x_2)\rangle &= \langle \Theta(x)\Theta(x_1)\rangle \del_{\nu} \delta^{(d)}(x-x_2) + \langle \Theta(x)\Theta(x_2)\rangle \del_\nu \delta^{(d)}(x-x_1) \, . 
\end{align}
A review of Ward identities as well as the derivation of these relations is in appendix \ref{ap:WardIdentities}.

In Lorentzian signature, variations of $\log Z$ produce time-ordered correlators.
In flat Minkowski space the time-ordered product is defined at separated points by  
\begin{align}\label{defT}
{\cal T} [{\cal O}_1(x_1) {\cal O}_2(x_2) \dotsi {\cal O}_n(x_n) ]
&= \sum_{\sigma \in S_n}  \theta(t_{\sigma_1} - t_{\sigma_2})\dotsi \theta(t_{\sigma_{n-1}} - t_{\sigma_n})  
{\cal O}_{\sigma_1}( x_{\sigma_1}) \dotsi {\cal O}_{\sigma_n}(x_{\sigma_n})\, ,
\end{align}
where the sum is over permutations of the set $\lbrace 1,\dots,n\rbrace$. 
The retarded product is defined at separated points by 
\begin{align}\label{defR}
\R[ {\cal O}(x); {\cal O}_1(x_1)\dotsi {\cal O}_{n-1}(x_{n-1})]
&=
(-i)^{n-1} \sum_{\sigma \in S_{n-1}} \theta(t-t_{\sigma_1})\dotsi\theta(t_{\sigma_{n-2}}-t_{\sigma_{n-1}})
\\
&\qquad  [\cdots [[{\cal O}(x), {\cal O}_{\sigma_1}(x_{\sigma_1})], {\cal O}_{\sigma_2}(x_{\sigma_2})] \dotsi {\cal O}_{\sigma_{n-1}}(x_{\sigma_{n-1}})] \notag\, ,
\end{align}
where the power of $i$ is chosen such that the retarded product is  Hermitian. 
In \eqref{defR}, the first argument is special, with ${\cal O}(x)$ always placed fully to the left in the nested commutator. The sum  symmetrizes over the remaining operators, $\mathcal{O}_1(x_1)\dots\mathcal{O}_{n-1}(x_{n-1})$. The retarded correlator vanishes unless $\mathcal{O}(x)$ is in the closed future lightcone of all the other operators.

Time-ordered and retarded correlators at coincident points are well defined as distributions, but their definition is not \eqref{defT}-\eqref{defR}. We will discuss the contact terms in section \ref{sec:anomaly}.

\subsection{Properties of the averaged null energy}

The averaged null energy (ANE) operator is the null energy integrated over a null ray,
\be 
\mathcal{E}_u(\v,\vec{x}) \equiv \int_{-\infty}^\infty \, du T_{uu}(u,\v,\vec{x})\, .\label{eq:ANECdef}
\ee
The ANE on a light-ray through the origin is denoted $\E_u(0) \equiv \E_u(0,\vec{0})$.

The ANE operator is a special case of a light-ray operator, which generally may or may not be given by the integral of a local operator. The systematic study of light-ray operators in conformal field theory in $d>2$ dimensions was initiated in \cite{Hofman:2008ar}. Light-ray operators appear in the lightcone OPE of local operators and thus control the high energy limit of correlation functions \cite{Hofman:2008ar,Hartman:2016lgu}. This was explored in the context of the Lorentzian inversion formula in \cite{Kravchuk:2018htv}. Light-ray operators also obey commutativity properties that lead to sum rules on the CFT data \cite{Kravchuk:2018htv,Kologlu:2019bco}. Their OPEs and crossing relations have been studied in 
\cite{Hofman:2009ug,Kologlu:2019mfz,Korchemsky:2019nzm,Chang:2020qpj,Chen:2022jhb,Chang:2022ryc}. Light-ray operators built from the stress tensor have special algebraic properties \cite{Cordova:2018ygx,Huang:2019fog,Huang:2020ycs,Belin:2020lsr,Besken:2020snx,Korchemsky:2021htm,Huang:2021hye,Huang:2022vcs,De:2022gjn} and encode some universal features of higher-dimensional CFT  \cite{Fitzpatrick:2019zqz,Fitzpatrick:2019efk,Fitzpatrick:2020yjb,Huang:2022vet}. They have been studied perturbatively in specific examples \cite{Bautista:2020bjy,Bautista:2022kmh,Caron-Huot:2022eqs}. Higher spin generalizations were studied in \cite{Hartman:2016lgu,Komargodski:2016gci,Meltzer:2018tnm,Kravchuk:2018htv}. For phenomenological applications, see e.g.~\cite{Korchemsky:1997sy,Sveshnikov:1995vi,Hofman:2009ug,Belitsky:2013ofa,Belitsky:2013bja,Belitsky:2013xxa,Henn:2019gkr,Kologlu:2019mfz,Gonzo:2020xza,Korchemsky:2021okt,Dixon:2019uzg,Chen:2020vvp}.

\subsubsection{Positivity}
Our convention for the stress tensor is such that the classical null energy condition is $\epsilon^\alpha \epsilon^\beta T_{\alpha\beta} \geq 0$ for $\epsilon^\alpha$ a forward-pointing null vector. The averaged null energy condition (ANEC) states that $\E_u$ is a positive semidefinite operator; that is,
\be 
\braket{\Psi|\mathcal{E}_u|\Psi}\geq 0 \, ,
\ee
in any state $\ket{\Psi}$. The ANEC has been established in various free theories \cite{Klinkhammer:1991ki,Wald:1991xn,Folacci:1992xg,Ford:1995gb} and more recently it was derived in interacting quantum field theories under quite general assumptions \cite{Faulkner:2016mzt,Hartman:2016lgu}. 

Non-negative operators that have vanishing vacuum expectation value must annihilate the vacuum $\ket{0}$, because otherwise a linear combination $|0\rangle + |\psi\rangle$ can be found in which the expectation value is negative \cite{Epstein:1965zza}. Thus the ANE operator satisfies 
\be 
\bra{0}\mathcal{E}_u =\mathcal{E}_u\ket{0} = 0\, .\label{eq:ANECanilvacuum}
\ee
This can also be shown by direct calculation: In a Wightman correlator $\langle 0 | \O_1(x_1)\dots \O_n(x) \E_u|0\rangle$, the $i\epsilon$ prescription for the $\O$ insertions ensures that the $u$-integral in $\int du T_{uu}$ can be smoothly deformed to vanish in the lower half plane \cite{Hartman:2016lgu}. In CFT, unitarity and representation theory of the conformal group imply that all light-ray operators annihilate the vacuum \cite{Kravchuk:2018htv}.

\subsubsection{The ANE in ordered correlators \label{sec:DifferentORderings}}

\begin{figure}[t] 
    \centering
    \includegraphics[width=0.95\linewidth]{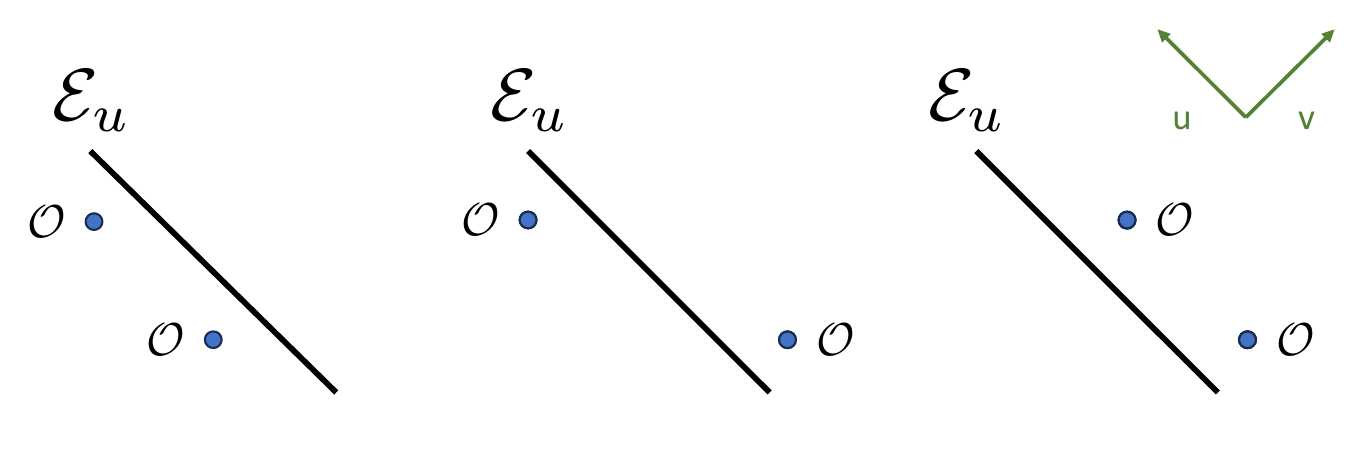}  
  \caption{Possible orderings of operators. As the ANE operator $\mathcal{E}_u$ annihilates the vacuum both on the right and on the left, 
  the cases where both operators $\mathcal{O}$ are on one or the other side of the ANE vanish. The only ordering that is not trivial is the one in the middle diagram, where there is one $\mathcal{O}$ operator on each side of the ANE operator.}
  \label{fig:figure1} 
\end{figure}

Time-ordered and retarded correlators of a light-ray operator are defined by ordering inside the integral,
\begin{align}
\langle {\cal T}[\E_u(v,\vec{x})\O(x_1)\O(x_2)]\rangle
&= \int_{-\infty}^{\infty} du\, \langle {\cal T}[T_{uu}(u,v,\vec{x}) \O(x_1)\O(x_2) ] \rangle \\
\langle {\cal R}[\E_u(v,\vec{x}); \O(x_1)\O(x_2)]\rangle
&= \int_{-\infty}^{\infty} du\, \langle 
\R[T_{uu}(u,v,\vec{x});  \O(x_1)\O(x_2) ] \rangle \ .
\end{align}
We will also use correlators where the ANE is in position space, but other operators are in momentum space. These are defined as
\begin{align}\label{orderedANEdef}
\langle {\cal T}[\E_u(v,\vec{x}) \O(k_1)\O(k_2) ]\rangle
&= \int d^d x_1 d^d x_2\, e^{ik_1\cdot x_1 + ik_2 \cdot x_2} \int du\, \langle {\cal T}[T_{uu}(u,v, \vec{x})\O(x_1)\O(x_2)]\rangle \\
\langle \R[\E_u(v,\vec{x}); \O(k_1)\O(k_2) ]\rangle
&= \int d^d x_1 d^d x_2\, e^{ik_1\cdot x_1 + ik_2 \cdot x_2} \int du\, \langle \R[T_{uu}(u,v, \vec{x}); \O(x_1)\O(x_2)]\rangle \ .  \notag
\end{align}
The fact that the ANE annihilates the vacuum can be used to simplify these correlators.
%
%
%
Consider the time-ordered product of the local null energy $T_{uu}$ and two local operators $\mathcal{O}_i \equiv \mathcal{O}(x_i)$. At separated points, 
\begin{align}
\mathcal{T}\left[T_{uu}(x)\mathcal{O}_1\mathcal{O}_2\right]&= \theta(t-t_1)\theta(t_1-t_2) T_{uu}\mathcal{O}_1\mathcal{O}_2 + \theta(t-t_2)\theta(t_2-t_1)T_{uu}\mathcal{O}_2\mathcal{O}_1\label{eq:220} \\
&+\theta(t_1-t)\theta(t-t_2)\mathcal{O}_1 T_{uu}\mathcal{O}_2+\theta(t_2-t)\theta(t-t_1)\mathcal{O}_2 T_{uu}\mathcal{O}_1\nonumber\\
&+\theta(t_1-t_2)\theta(t_2-t)\mathcal{O}_1 \mathcal{O}_2T_{uu}+\theta(t_2-t_1)\theta(t_1-t)\mathcal{O}_2 \mathcal{O}_1T_{uu}\, .\nonumber\
\end{align}
Using \eqref{eq:ANECanilvacuum} and causality, which implies that commutators must vanish for spacelike separated operators, it is clear that if we integrate the stress tensor over a null ray, only two of these six orderings are non-vanishing in the 3-point function. In particular, only the second line of \eqref{eq:220} contributes. This is pictured on figure \ref{fig:figure1}. Placing the ANE through the origin, we obtain
\begin{align}
\braket{\mathcal{T}[\mathcal{E}_u(0)\mathcal{O}_1\mathcal{O}_2]}&= \theta(v_1)\theta(-v_2)\braket{\mathcal{O}_1 \mathcal{E}_u(0)\mathcal{O}_2}+\theta(v_2)\theta(-v_1)\braket{\mathcal{O}_2 \mathcal{E}_u(0)\mathcal{O}_1}\\
&\quad + \mbox{contact terms} \notag
\end{align}
where we used causality to write the step functions in terms of the null coordinate $v$. With an $i\epsilon$-prescription appropriate to the operator ordering, the light-ray integral is convergent so it can be performed safely.
The retarded correlator also simplifies. At separated points,
\begin{align}
\R\left[T_{uu}(0);\mathcal{O}_1\mathcal{O}_2\right] =& -\theta(-t_1)\theta(t_{1}-t_{2})\left[\left[T_{uu}(0),\mathcal{O}_1\right], \mathcal{O}_2\right]-\theta(-t_2)\theta(t_{2}-t_{1})\left[\left[T_{uu}(0),\mathcal{O}_2\right], \mathcal{O}_1\right]\, .\label{eq:RetardFootnote2}
\end{align}
Expanding the nested commutators in \eqref{eq:RetardFootnote2}, integrating the stress tensor to obtain an ANE operator, and using the fact that the ANE annihilates the vacuum to remove all the correlation functions where the ANE is either all the way to the right or to the left, we obtain 
%
\begin{align}
\braket{\R[\mathcal{E}_u(0);\mathcal{O}_1\mathcal{O}_2]} &=\theta(-v_1)\theta(-v_2)\left[\braket{\mathcal{O}_1\mathcal{E}_u(0)\mathcal{O}_2}+\braket{\mathcal{O}_2\mathcal{E}_u(0)\mathcal{O}_1}\right] \label{eq:Retarded2d1bb}\\
&\quad + \mbox{contact terms} \notag .
\end{align}
We can then transform the local $\O_i$ operators to momentum space, such that ordered correlators of the ANE  operator can be expressed in terms of Wightman functions as
\begin{align}
\langle \R[\E_u(v,\vec{x}); \O(k_1)\O(k_2) ]\rangle
&= 2 \int_{v_1<v} d^d x_1 \int_{v_2<v} d^d x_2\,  
e^{ik_1\cdot x_1 + ik_2 \cdot x_2} \langle \O(x_1)\E_u(v,\vec{x})\O(x_2)\rangle_{\sep}
\notag \\
&\quad + \mbox{contact terms} \ .
\end{align}
and
\begin{align}
\langle {\cal T}[\E_u(v,\vec{x}) \O(k_1)\O(k_2) ]\rangle
&= 2 \int_{v_1>v} d^d x_1 \int_{v_2<v} d^d x_2 
\, e^{ik_1\cdot x_1 + ik_2 \cdot x_2} \langle \O(x_1)\E_u(v,\vec{x})\O(x_2)\rangle_{\sep}
\notag \\
&\quad + \mbox{contact terms} \ .
\end{align}
Wightman functions do not have contact terms, but for clarity we have written `sep' for `separated' to indicate that there are no contact terms included in the first lines.

\subsubsection{Response function}
Retarded correlation functions measure the response of an operator to a deformation of the Hamiltonian. Turning on a time-dependent deformation $H_0 \to H_0 + V(t)$, the expectation value of an operator $A$ in the interaction picture is
\begin{align}
\langle A(x) \rangle_{\rm def} = \langle U(t)^{\dagger} A(x) U(t) \rangle \ , \qquad U(t) = 
{\cal T}\!\exp\left(- i \int^t dt' V(t') \right)\, ,
\end{align}
with the right-hand side evaluated in the undeformed theory. Expanding perturbatively in $V$ gives the Dyson series of retarded correlation functions. For a deformation by a local operator such that $\delta S = \int J \O$ with real $J$, the first few terms are\footnote{These are full correlators. Elsewhere we use $\langle \cdot \rangle$ for connected correlators. Once we choose $A = \E_u$ there is no difference for the low-order terms written, because the disconnected terms vanish using $\E_u|0\rangle=0$.}
\begin{align}
\langle A(x)\rangle_{\rm def}  &= \langle A(x)\rangle
 + \int d^d x_1\, J(x_1) \langle \R[A(x); \O(x_1)]\rangle\\
 &\quad 
 + \frac{1}{2} \int d^d x_1\, d^d x_2\, J(x_1)J(x_2)\langle \R[A(x); \O(x_1)\O(x_2)]\rangle + \cdots\notag
\end{align}
This provides a physical interpretation for the retarded correlation functions of the ANE. Let us insert the ANE through the origin, and turn off the deformation at $v=-\epsilon$, slightly before  the ANE insertion, to avoid contact terms. The first and second order terms vanish using $\E_u|0\rangle = 0$, so the leading term is
\begin{align}\label{edyson}
\langle \E_u(0)\rangle_{\rm def} &=
\langle \Psi | \E_u(0) | \Psi\rangle\\
&= \frac{1}{2} \int_{v_1<0} d^d x_1 \int_{v_2<0} d^d x_2\, J(x_1)J(x_2)\langle \R[\E_u(0); \O(x_1)\O(x_2)]\rangle_{\rm sep}\notag\, ,
\end{align}
where the state is $|\Psi\rangle = U(v=-\epsilon)|0\rangle$ and in the second line we have sent $\epsilon \to 0$.
In particular, if we take $\O$ to be the trace of the stress tensor, then the correlator $\langle \R[\E_u; \Theta\Theta]\rangle$ that will play a central role in this paper is the expectation value $\langle \E_u\rangle_{\rm def}$ in a Weyl-deformed theory. Since \eqref{edyson} is an expectation value, it is required to be nonnegative by the ANEC in the original theory on a flat metric.\footnote{
Note that there are two different things one could mean by the averaged null energy in a Weyl-deformed theory. Here we mean $\int du T_{uu}$ in the theory with metric $e^{\omega}(-dudv+d\vec{x}^2)$. Some papers on the ANEC in curved spacetime study instead the quantity $\int d\lambda T_{\lambda\lambda} = \int du e^{-\omega} T_{uu}$ where $\lambda$ is an affine parameter in this metric \cite{Visser:1994jb,Urban:2009yt}. These papers show that the ANEC (in the latter sense) can be violated in quantum field theory on a conformally flat background. The ANEC is believed to hold in gravitational theories, but only if the null ray is achronal  and the background satisfies the Einstein equations \cite{Graham_2007}.}

\section{Contact terms in Lorentzian correlators}\label{sec:anomaly}


In this section we discuss general properties of contact terms in Lorentzian correlation functions,  then use the trace anomaly to calculate the contact terms in $\langle \T[T_{\alpha\beta}\Theta\Theta]\rangle$ and $\langle \R[T_{\alpha\beta}; \Theta\Theta]\rangle$ in four-dimensional CFT. From the latter results we write the Euler anomaly $a$ in CFT in terms of the averaged null energy.

\subsection{General properties}\label{ss:contactproperties}
The conformal anomaly produces contact terms in correlation functions of the stress tensor. These contact terms are most commonly discussed in Euclidean signature, where they appear as Dirac $\delta$-functions in position space, or polynomials in momentum space. In Lorentzian signature, contact terms appear in the time-ordered, anti-time-ordered, retarded, and advanced momentum-space correlators. These four types of correlators are all related to each other by analytic continuation in momenta. This implies that, in fact, all the different correlators in momentum space share the exact same contact terms up to phases. The goal of this section is to explain this in detail. 



The definitions of the different orderings of operators \eqref{defT} and \eqref{defR} only apply at separated points.  Correlators at coincident points, i.e.~contact terms, are defined by consistency relations or (equivalently) by variations as in \eqref{eq:Tcorrelator1}. In a CFT in even dimensions, the conformal anomaly signals the fact that the stress tensor cannot be simultaneously conserved 
and traceless at coincident points.
Because we defined the stress tensor by the metric variation, as in equation \eqref{eq:Tcorrelator1}, we have opted for conservation over tracelessness, and the trace anomaly leads to coincident point contributions to correlation functions. 

In Lorentzian signature, it is easier to study contact terms in momentum space.  Let us first review the analytic structure of momentum-space correlators (see \cite{Meltzer:2021bmb} for a more comprehensive review).
%
%
In a unitary quantum field theory, physical states are required to have momentum in the closed forward cone,  $\overline{V}_+ = \{q| q^2 \leq 0 , q^t \geq 0\}$. This is one of the Wightman axioms (see e.g.~\cite{Haag:1992hx}).\footnote{We are also assuming the existence of the time-ordered product, which is an extra assumption that has not been derived from the Wightman axioms. The $\R$-product is then defined from the $\T$-product. See e.g.~\cite{Kravchuk:2021kwe,Meltzer:2021bmb}.}  It follows that $\O(q)|0\rangle = 0$ for $q \notin\overline{V}_+$. Moreoever,
\be 
\mathcal{T}\left[\mathcal{O}(k_1)\dots\mathcal{O}(k_n)\right]\ket{0}=\R\left[\mathcal{O}(k_1); \dots\mathcal{O}(k_n)\right]\ket{0}=0\, ,\qquad \text{ if} \qquad \sum_{i=1}^n k_i \not \in \bar{V}_{+} \ .
\ee
Similarly, a string of operators, in any ordering, vanishes when acting on the left vacuum $\bra{0}$ if the sum of their momenta is outside the closed backward cone, $\overline{V}_- = \{q| q^2 \leq 0 , q^t \leq 0\}$.
%
%
%
%
These properties can be used to show that if all external and internal momenta lie outside the region $\overline{V}_-$, 
then the retarded and time-ordered correlators are equal up to a phase:
\begin{align}\label{rtaugeneral}
\lll \R[\O_n(k_n); O_1(k_1) \dots \O_{n-1}(k_{n-1}) \rr
= (-i)^{n-1} \lll
{\cal T}[ \O_1(k_1) \dots \O_n(k_n)] \rr
 \, , \qquad 
 \sum_{i \in I} k_i \notin \overline{V}_- \, ,\end{align}
for all $ I \subseteq\{1,\dots,n-1\}$. 
This implies that the ${\cal T}$ and $\R$ momentum-space correlators are analytic continuations of the same function, up to the phase $(-i)^{n-1}$. (The advanced and anti-time-ordered functions are also analytic continuations of this same function, but we will not need them.) There are poles or branch cuts so that the correlators differ for momenta in $\overline{V}_-$.

The general derivation of \eqref{rtaugeneral} is reviewed in e.g.~\cite{Meltzer:2021bmb}. We will sketch it briefly for the 2-point and 3-point functions. For the 2-point function, the orderings are trivially related by
\begin{align}\label{RT2}
i\R[\O_2(x_2); &\O_1(x_1)] - {\cal T}[ \O_1(x_1)\O_2(x_2) ]\\
&= \theta(t_2-t_1)[\O_2(x_2), \O_1(x_1)]  - \theta(t_1-t_2)\O_1(x_1)\O_2(x_2) - \theta(t_2-t_1)\O_2(x_2)\O_1(x_1)\notag \\
&= -\O_1(x_1)\O_2(x_2)\notag\, .
\end{align}
The second line in this expression makes sense only at separated points, but the first and third line are equal as distributions. The Fourier transform of the Wightman function $\langle \O_1(x_1)\O_2(x_2)\rangle$ vanishes for $k_2 \notin \overline{V}_+$, because $\O(k_2)|0\rangle$ must create a state in the physical spectrum to obtain something non-vanishing.  Therefore applying this to \eqref{RT2} we have $\lll \R[ \O_2(-k) ; \O_1(k)\rr = - i \lll \O_1(k)\O_2(-k)\rr$ for $k \notin \overline{V}_-$. This establishes \eqref{rtaugeneral} for 2-point functions. For the 3-point function, the starting point is the identity
\begin{align}
\R[\O_3; \O_1\O_2] + {\cal T}[ \O_1 \O_2 \O_3] = i \O_2 \R[\O_3; \O_1] + i \O_1 \R[\O_3; \O_2] + i \R[\O_1; \O_2] \O_3 + \O_2 \O_1 \O_3
\end{align}
with $\O_i \equiv \O_i(x_i)$. If $\lbrace k_1,\, k_2,\, k_1+k_2\rbrace \notin \overline{V}_-$ then the Fourier transform of the right-hand side vanishes, since $\langle 0|\O_1(k_1) =\langle 0|\O_2(k_2) =0$, and $\O_3(-k_1-k_2)|0\rangle =0$. This implies $\R[\O_3; \O_1\O_2] =- {\cal T}[ \O_1 \O_2 \O_3]$, as stated in  \eqref{rtaugeneral}.

We now turn to contact terms. By definition, contact terms are analytic in momenta. Therefore \eqref{rtaugeneral} holds everywhere without any restriction on momenta,
\begin{align}\label{econt}
\lll \R[\O_n(k_n); O_1(k_1) \dots \O_{n-1}(k_{n-1}) \rr_{\rm contact}
= (-i)^{n-1} \lll
{\cal T}[ \O_1(k_1) \dots \O_n(k_n)] \rr_{\rm contact}\, ,
 \quad \forall k_i\  . 
\end{align}
There can be partial contact terms which are analytic in some momenta but not others; \eqref{econt} applies to the full contact terms, analytic in all momenta.

Euclidean momentum-space correlators are related to time-ordered Lorentzian correlators by a Wick rotation. Thus we can summarize the situation as follows. Up to simple phase factors, exactly the same contact terms appear as polynomials in momenta in the Euclidean, time-ordered, anti-time-ordered, retarded, and advanced correlators. Wightman functions do not have contact terms, because an analytic function that vanishes for spacelike momenta is identically zero. 

It is important to keep in mind that whereas retarded and time-ordered correlators are well defined distributions, the expansions \eqref{defT} can only be used at separated points. Contact terms in Euclidean correlators come from imposing conservation at coincidence points in the Euclidean path integral. Similarly, contact terms in time-ordered correlators come from imposing conservation at coincident points in the Lorentzian path integral, and contact terms in retarded correlators can be understood from the path integral on a Schwinger-Keldysh contour.

\subsection{Contact terms in four-dimensional CFT}\label{ss:cftLocalContact}
In a four-dimensional CFT in curved spacetime, the trace of the stress tensor is
\begin{align}\label{theta4d}
\langle \Theta\rangle&= -a E_4 + c W_{\mu\nu\rho\sigma}^2 + b_1 \Box R + b_2 \Lambda^2 R + b_3 \Lambda^4 \ ,
\end{align}
where $E_4$ is the Euler density and $W_{\mu\nu\rho\sigma}$ is the Weyl tensor (see appendix \ref{ap:FourDim}) \cite{Deser:1976yx,Duff:1977ay} (see also the reviews \cite{Duff:1993wm,Nakayama:2013is}). The Weyl anomaly consists of the first two terms, with coefficients $a$ and $c$. The other terms, where $\Lambda$ is the UV cutoff and the $b_i$ are dimensionless coefficients, can be removed by local counterterms, but if they are set to zero in the UV, they will be generated along an RG flow.\footnote{ The corresponding terms in the Euclidean effective action are
\begin{align}
\log Z_b &= \int d^4 x \sqrt{g} \left( \frac{b_3}{4} \Lambda^4 + \frac{ b_2}{2} \Lambda^2 R
 + b_1' R^2 + b_1'' R_{\mu\nu}R^{\mu\nu} \right)\, ,
\end{align}
such that $b_1 \Box R + b_2 \Lambda^2 R + b_3 \Lambda^4 = \frac{2}{\sqrt{g}} \frac{\delta}{\delta \omega} \log Z_b$
with $b_1 = -12b_1'-4b_1''$ where we used the Bianchi identity. }
We retain them in order to ensure that the eventual sum rule is independent of non-universal terms.

Dimensional analysis also allows a contribution to the trace proportional to $R^2$, but this term is eliminated by the Wess-Zumino consistency condition which requires $\langle\Theta(x_1)\Theta(x_2)\rangle$ to be symmetric \cite{Wess:1971yu}. In CP-violating theories there could also be a Pontryagin term $ \epsilon^{\mu\nu\rho\sigma} R_{\mu\nu}{}^{\alpha\beta} R_{\alpha\beta \rho\sigma}$. We will not keep track of it, but it does not contribute to $\langle \Theta\Theta T_{uu}\rangle $ so it does not affect the conclusions.

%

\subsubsection*{2-point function $\langle\Theta\Theta\rangle$}

Correlation functions of the stress tensor involving at least one insertion of $\Theta$ can be calculated by varying \eqref{theta4d}. For example, the trace 2-point function in Euclidean signature is
\begin{align}
\langle\Theta(x_1)\Theta(x_2)\rangle &= 
-\frac{2}{\sqrt{g(x_1)} \sqrt{g(x_2)}}  g^{\alpha\beta}(x_2) \frac{\delta }{\delta g^{\alpha\beta}(x_2)}\left( \sqrt{g(x_1)} \langle\Theta(x_1)\rangle \right)  \ ,
\end{align}
with the metric implicitly set to $g_{\mu\nu}=\delta_{\mu\nu}$ at the end. Varying \eqref{theta4d} and then transforming to momentum space (this computation can be found in appendix \ref{ap:FourDim}) gives the Euclidean 2-point function
\begin{align}
\lll \Theta(\K) \Theta(-\K) \rr &=  - 6 b_1 \K^4 + 6 b_2 \Lambda^2 \K^2 + 4 b_3 \Lambda^4 \ .
\end{align}
The Wick rotation to Lorentzian signature with $(\K^\tau, \K^i) = (i k^t, k^i)$  produces the time-ordered correlator
\begin{align}
\lll \T[ \Theta(k)\Theta(-k) ] \rr &= -i( - 6 b_1 k^4 + 6 b_2 \Lambda^2 k^2 + 4 b_3 \Lambda^4 ) \ .
\end{align}
This is a pure contact term, so the retarded correlator is equal up to a phase, as discussed in section \ref{ss:contactproperties}. This amounts to 
\be 
\lll \R[ \Theta(k); \Theta(-k) ] \rr = -i \lll \T[ \Theta(k)\Theta(-k) ] \rr =  6 b_1 k^4 - 6 b_2 \Lambda^2 k^2 - 4 b_3 \Lambda^4 \, .
\ee
Note that the anomaly coefficients $(a,\, c)$ do not appear in $\langle\Theta\Theta\rangle$. The 2-point function $\langle T_{\alpha\beta}\Theta\rangle$ is calculated in appendix \ref{ap:FourDim}, and is also independent of $a$ or $c$. To study the anomaly, we therefore proceed to higher variations.

\subsubsection*{3-point function $\langle T_{uu} \Theta\Theta\rangle$}

Consider $\langle T_{uu}\Theta\Theta\rangle$, with $T_{uu}$ the null energy. To calculate the Lorentzian correlators, we vary the trace \eqref{theta4d}, transform to momentum space, and Wick rotate. The calculation is a straightforward application of the definitions in section \ref{sec:DefCor} and is detailed in appendix \ref{ap:FourDim}. The results are
\begin{align}\label{R34d}
\lll \R[T_{uu}(k_3) ; \Theta(k_1)\Theta(k_2)]\rr &= 8a(k_{1u}^2 k_2^2 + k_{2u}^2 k_1^2 - 2 k_{1u}k_{2u}\,k_1\sdot k_2)\\
&\quad  +4b_1[(k_{1u}+k_{2u})^2k_1\sdot k_2-3 k_{1u}k_{2u}(k_1^2+k_2^2)]\notag\\
&\quad  - 4 b_2\Lambda^2(k_{1u}^2+k_{2u}^2 - k_{1u}k_{2u})\notag\, ,
\end{align}
and
\begin{align}\label{R34dT}
 \lll {\cal T}[\Theta(k_1)\Theta(k_2)T_{uu}(k_3)]\rr &=
 - \lll \R[T_{uu}(k_3); \Theta(k_1)\Theta(k_2)]\rr \ .
\end{align}
The Weyl-squared anomaly $c$ dropped out, but the Euler anomaly $a$ appears in the first line. In a given CFT, the coefficients $b_1$ and $b_2$ can always be set to zero by local counterterms, but they are generated by RG flows so they cannot be set to zero simultaneously in the UV and IR.

%
%

%
%


\subsection{The trace anomaly from averaged null energy}
Contact terms in correlation functions involving light-ray operators are derived by integrating the local correlators. It is convenient to work in momentum space. We will now use \eqref{R34d} to calculate the contact terms in $\langle \E_u \Theta\Theta\rangle$ in a 4d CFT, and derive a universal formula for the $a$ anomaly in terms of this correlator. 

The ANE operator through the origin is 
\begin{align}
\E_u(0) = \int du\, T_{uu}(u,v=0,\vec{x}=0)  =  \int \frac{dk_v\, d^2 \vec{k}}{\pi(2\pi)^2} T_{uu}(k_u=0,k_v, \vec{k})\, .
\end{align}
Ordered correlators of the ANE are defined in \eqref{orderedANEdef}.
By setting $k_{3u} = 0$ and integrating \eqref{R34d} over the other components of $k_3$, we find the retarded correlator
\begin{align}\label{answerE4d}
\langle \R[\E_u(0);\Theta(k_1)\Theta(k_2)\rangle
&= 8\pi k_{1u}^2 [2a (k_1+k_2)^2 + 3b_1 (k_1^2+k_2^2) - 3 b_2\Lambda^2] \delta(k_{1u}+k_{2u}) \, ,
\end{align} 
and the time-ordered correlator
\begin{align}
\langle \T[\E_u(0)\Theta(k_1)\Theta(k_2)\rangle
&=
-\langle \R[\E_u(0);\Theta(k_1)\Theta(k_2)\rangle \ .
\end{align}
Note that the argument of $\E_u(0)$ is always a position, $(v,\vec{x}) = (0, \vec{0})$, not a momentum. 

The  equation \eqref{answerE4d} can be inverted to solve for the three trace coefficients $a,b_1,b_2$ by taking advantage of the different transverse momentum structures. After setting $k_{1v} = k_{2v} = 0$, the momentum dependence of the Euler anomaly term is $(k_1+k_2)^2 \to (\vec{k}_1 + \vec{k}_2)^2$. This is the only term involving $\vec{k}_1 \cdot \vec{k}_2$, so to extract the $a$ coefficient we integrate \eqref{answerE4d} over $k_{2u}$, act with $\partial_{k_{1u}}^2\vec{\p}_{k_1} \sdot \vec{\p}_{k_2}$, then set all momenta to zero. The result is a universal formula for the Euler term in the conformal anomaly in a 4d CFT:
\begin{align}\label{aCFT}
a &= \frac{1}{64}\int d^4 x_1\, d^4 x_2\, u_1^2 \delta(u_2)\, \vec{x}_1 \sdot \vec{x}_2
\langle \R[\mathcal{E}_u(0); \Theta(x_1)\Theta(x_2)]\rangle \, .
\end{align}
This is the key relation that we will use in the derivation of the sum rule. 
By translation invariance we can also fix the $u$-position of $T_{uu}$ instead of one of the traces, giving an expression in terms of the local null energy,
\begin{align}\label{aCFTlocal}
a = \frac{1}{64} \int d^4 x_1 \, d^4 x_2 \,(u_1-u_2)^2 \,\vec{x}_1 \cdot \vec{x}_2 \langle \R[T_{uu}(0);\Theta(x_1)\Theta(x_2)]\rangle \ .
\end{align}
%
The non-universal terms can be extracted from \eqref{answerE4d} by acting with different momentum derivatives. Choosing the differential operator $\p_{k_{1u}}^2$ gives
\begin{align}
b_2\Lambda^2 = \frac{1}{24}\int d^4x_1\, d^4x_2 \, u_1^2 \,\delta(u_2) \langle \R[\E_u(0); \Theta(x_1)\Theta(x_2)]\rangle \, ,
\end{align}
and the other coefficient can be isolated using $\p_{k_{1u}}^2(\vec{\p}_{k_1} - \vec{\p}_{k_2})^2$, such that
\begin{align}
b_1 &= \frac{1}{192} 
\int d^4 x_1\, d^4 x_2\, u_1^2\, \delta(u_2) (\vec{x}_1-\vec{x}_2)^2
\langle \R[\E_u(0); \Theta(x_1)\Theta(x_2)]\rangle \, .
\end{align}

\section{The $a$-theorem from the ANEC}\label{sec:atheorem}



\begin{figure}
\begin{center}

\tikzset{every picture/.style={line width=0.75pt}} 

\begin{tikzpicture}[x=0.75pt,y=0.75pt,yscale=-1,xscale=1]

\draw [color={rgb, 255:red, 74; green, 144; blue, 226 }  ,draw opacity=1 ][line width=2.25]    (117.5,287.53) -- (117.5,50.89) ;
\draw [shift={(117.5,45.89)}, rotate = 90] [fill={rgb, 255:red, 74; green, 144; blue, 226 }  ,fill opacity=1 ][line width=0.08]  [draw opacity=0] (14.29,-6.86) -- (0,0) -- (14.29,6.86) -- cycle    ;
\draw [color={rgb, 255:red, 2; green, 57; blue, 123 }  ,draw opacity=1 ][fill={rgb, 255:red, 80; green, 116; blue, 159 }  ,fill opacity=1 ][line width=1.5]    (235,161.09) .. controls (236.67,162.76) and (236.67,164.42) .. (235,166.09) .. controls (233.33,167.76) and (233.33,169.42) .. (235,171.09) .. controls (236.67,172.76) and (236.67,174.42) .. (235,176.09) .. controls (233.33,177.76) and (233.33,179.42) .. (235,181.09) .. controls (236.67,182.76) and (236.67,184.42) .. (235,186.09) .. controls (233.33,187.76) and (233.33,189.42) .. (235,191.09) .. controls (236.67,192.76) and (236.67,194.42) .. (235,196.09) .. controls (233.33,197.76) and (233.33,199.42) .. (235,201.09) .. controls (236.67,202.76) and (236.67,204.42) .. (235,206.09) .. controls (233.33,207.76) and (233.33,209.42) .. (235,211.09) .. controls (236.67,212.76) and (236.67,214.42) .. (235,216.09) .. controls (233.33,217.76) and (233.33,219.42) .. (235,221.09) .. controls (236.67,222.76) and (236.67,224.42) .. (235,226.09) -- (235,227.31) -- (235,235.31) ;
\draw [shift={(235,239.31)}, rotate = 270] [fill={rgb, 255:red, 2; green, 57; blue, 123 }  ,fill opacity=1 ][line width=0.08]  [draw opacity=0] (11.61,-5.58) -- (0,0) -- (11.61,5.58) -- cycle    ;

\draw (92,291.75) node [anchor=north west][inner sep=0.75pt]  [color={rgb, 255:red, 74; green, 144; blue, 226 }  ,opacity=1 ] [align=left] {Energy};
\draw (17,61.63) node [anchor=north west][inner sep=0.75pt]    {$CFT_{UV}$};
\draw (17,249.79) node [anchor=north west][inner sep=0.75pt]    {$CFT_{IR}$};
\draw (17.5,136.82) node [anchor=north west][inner sep=0.75pt]    {$QFT$};
\draw (185.5,134.36) node [anchor=north west][inner sep=0.75pt]    {$\langle \R[\mathcal{E}_{u} ;\Theta \Theta ] \rangle \ $};
\draw (185,61.56) node [anchor=north west][inner sep=0.75pt]    {$\langle \R[\mathcal{E}_{u} ;\Theta \Theta ] \rangle _{UV} \ $};
\draw (185.5,247.28) node [anchor=north west][inner sep=0.75pt]    {$\langle \R[\mathcal{E}_{u} ;\Theta \Theta ] \rangle _{IR} \ \ $};
\draw (266,188.3) node [anchor=north west][inner sep=0.75pt]  [color={rgb, 255:red, 2; green, 57; blue, 123 }  ,opacity=1 ] [align=left] {\textcolor[rgb]{0.01,0.22,0.48}{Low momentum limit }};

\end{tikzpicture}
\end{center}
\caption{\small
Logic of the anomaly matching sum rule. We consider the correlator 
$\langle \R[\E_u; \Theta\Theta]\rangle$ in a QFT that flows between two fixed points. At low momentum it must reproduce the IR anomaly. This can be viewed as a sum rule, with contributions from the UV and from separated points along the flow. 
\label{fig:rgflows}
}
\end{figure}
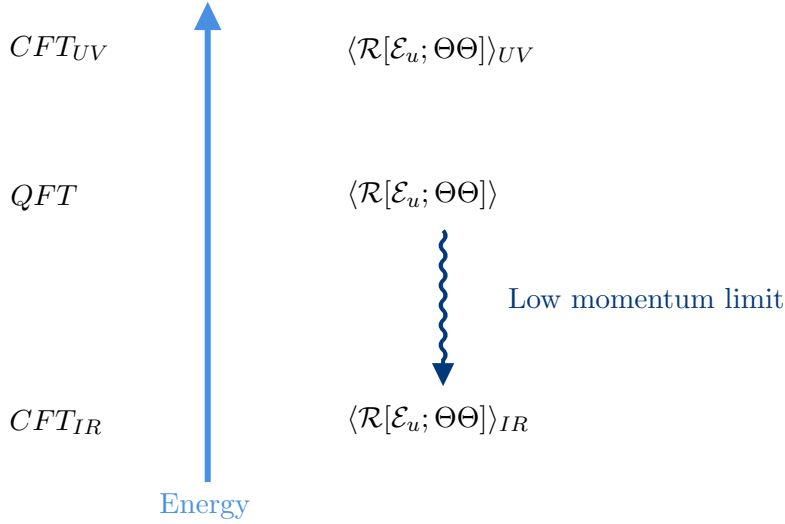

We now turn to the study of non-conformal quantum field theories and derive the sum rule relating $\langle \Theta \E_u \Theta\rangle$ to the change in the Euler anomaly, $\Delta a= a_{UV} - a_{IR}$, along an RG flow. The strategy is sketched in figure \ref{fig:rgflows}. We will apply the CFT relation \eqref{aCFT} to the infrared theory, and then decompose the right-hand side into contributions from the UV fixed point and contributions along the RG flow.

\subsection{Matching in the infrared}

Consider a four-dimensional QFT that flows from CFT$_{\rm UV}$ at high energies to CFT$_{\rm IR}$ at low energies. For a QFT with a mass scale $M$, high energies means $E \gg M$ and low energies means $E \ll M$. Throughout this section and for the rest of the paper, we use brackets $\langle \cdot \rangle$ to denote connected correlation functions in the QFT, with $\langle \cdot \rangle_{\rm UV}$ and $\langle \cdot \rangle_{\rm IR}$ used for connected correlators in the CFTs at the fixed points. Let us choose the counterterms to set $b_1=b_2=b_3=0$ in the UV CFT. The mass scale $M$ plays the role of the UV cutoff in the IR theory, so
\begin{align}\label{thetaUVIR}
\langle \Theta\rangle_{{\rm UV}} &= -a_{UV} E_4 + c_{UV} W_{\mu\nu\alpha\beta}^2\, , \\
\langle \Theta\rangle_{{\rm IR}} &= -a _{IR}E_4+c_{IR} W_{\mu\nu\alpha\beta}^2 + b_1 \Box R + b_2 M^2 R + b_3 M^4 \ .\notag
\end{align}
The renormalization group relates the correlation functions of the QFT to those of the CFTs. In the infrared, the statement that the QFT flows to CFT$_{\rm IR}$  means that 
correlation functions of the QFT approach those of CFT$_{\rm IR}$ when all momentum invariants are small compared to the mass, i.e.  $\K_i \cdot \K_j \ll M^2$. 
Correlators of the QFT involving $\Theta$ must agree with the above expressions for $\langle \Theta\rangle_{{\rm IR}}$ at low momenta. In particular, the 3-point function satisfies
\begin{align}\label{tttIR}
 \lll  T_{\alpha\beta}(\K_3)\Theta(\K_1)\Theta(\K_2)\rr
&\approx 
\lll T_{\alpha\beta}(\K_3)\Theta(\K_1)\Theta(\K_2)\rr_{IR}\, ,
\quad \mbox{for} \quad
\frac{\K_i \sdot \K_j}{M^2} \ll 1\, ,
\end{align}
where the right-hand side is  the CFT correlator obtained in \eqref{R34d} (in Lorentzian) or \eqref{eq:Tuvthetatheta4dap} (in Euclidean), which is a pure contact term that is analytic in all momenta. Corrections to \eqref{tttIR} are suppressed by positive powers of the invariants $\frac{\K_1^2}{M^2},\, \frac{\K_2^2}{M^2},\, \frac{\K_1\cdot \K_2}{M^2}$, multiplied by logarithms. 
The same statements apply to the time-ordered and retarded correlators in Lorentzian signature. In the ultraviolet, the statement that the QFT flows to CFT$_{\rm UV}$ at high energies means that the correlators of the two theories agree up to corrections suppressed by the invariants $\frac{M^2}{\K_i \cdot \K_j}$.


In order for \eqref{tttIR} to hold, it is important to keep all terms in the IR  trace  in \eqref{thetaUVIR}, including those multiplying the UV cutoff. Usually in the discussion of a CFT in isolation the $b$-coefficients are renormalized to zero.  These terms, unlike the anomaly coefficients, are not  intrinsic to the IR CFT, but they can be calculated from the QFT using \eqref{tttIR}.

\subsection{Derivation of the sum rule}\label{ss:deriveAsum}
The IR relation \eqref{tttIR} can be viewed as a sum rule that relates the UV to the IR.  All of the terms in the CFT correlator must match the low-momentum expansion of the QFT. In four dimensions, this means all terms up to and including $O(\K^4)$. Writing out the Fourier transform, we have
\begin{align}\label{matchWithErr}
\int d^4 x_1 d^4 x_2 d^4 x_3\, e^{i\K_1\cdot x_1 + i \K_2 \cdot x_2 + i \K_3 \cdot x_3}\langle T_{\alpha\beta}(x_3)\Theta(x_1)\Theta(x_2)\rangle 
&\approx \langle T_{\alpha\beta}(\K_3) \Theta(\K_1) \Theta(\K_2)\rangle_{IR} \, ,
\end{align}
at low momentum. 
The Fourier integral has three types of contributions, depending on how many points are coincident: separated terms with no delta functions, `partial contact' (PC) terms with a single delta function, and full contact terms with two delta functions. The full contact terms are controlled by the UV fixed point, while the partial contact terms depend on both the UV CFT and the RG flow. 
Let us write these three terms as
\begin{align}\label{tsplit}
\langle T_{\alpha\beta}(x_3)\Theta(x_1)\Theta(x_2)\rangle
&= 
\langle T_{\alpha\beta}(x_3)\Theta(x_1)\Theta(x_2)\rangle_\sep
 + \langle T_{\alpha\beta}(x_3)\Theta(x_1)\Theta(x_2)\rangle_{PC}\\
 &\qquad 
 +
\langle T_{\alpha\beta}(x_3)\Theta(x_1)\Theta(x_2)\rangle_{UV}
   \ ,\notag
\end{align}
which have zero, one, and two delta functions respectively.  The separated term $\langle \cdot \rangle_\sep$ is by definition a non-singular distribution that agrees with the full correlator $\langle \cdot \rangle$ when no two points are coincident, and integrates to zero against test functions with support in a vanishingly small region; in other words, it is the correlator with all contact terms discarded. For the split in \eqref{tsplit} to be well defined, each term must individually be a well defined distribution. Equivalently, the split only makes sense when integrated test functions such that all three integrals converge separately. 

We apply the split \eqref{tsplit} and the matching condition \eqref{matchWithErr} to the ANE correlator. Isolating the contribution from separated points, we have
\begin{align}
\int d^4 x_1 d^4 x_2 \, &e^{ik_1\cdot x_1 + i k_2 \cdot x_2 }
\langle \R[\E_u(0); \Theta(x_1)\Theta(x_2)]\rangle_\sep +PC \\
&\approx \langle \R[\E_u(0); \Theta(k_1)\Theta(k_2)]\rangle_{IR}
 - \langle \R[\E_u(0); \Theta(k_1)\Theta(k_2)]\rangle_{UV}\, , \notag
\end{align}
where $PC$ stands for the term originating from the partial contact terms:
\begin{align}
PC &= \int d^4 x_1 d^4 x_2\,  e^{ik_1\cdot x_1 + i k_2 \cdot x_2 }
\langle \R[\E_u(0); \Theta(x_1)\Theta(x_2)]\rangle_{PC}\, .
\end{align}
Plugging in the UV and IR CFT correlators obtained in \eqref{answerE4d}, and recalling that we tuned the non-universal terms to zero in the UV, gives
\begin{align}
\int d^4 x_1 d^4 x_2\,  &e^{ik_1\cdot x_1 + i k_2 \cdot x_2 }
\langle \R[\E_u(0); \Theta(x_1)\Theta(x_2)\rangle_\sep + PC \\
&\approx
16\pi k_{1u}^2\Big[ (a_{IR} - a_{UV})(k_1+k_2)^2  - \tfrac{3}{2} b_1(k_1^2+k_2^2) - \tfrac{3}{2}b_2 M^2 \Big] \delta(k_{1u} + k_{2u})
 \notag \ .
\end{align}
Next, we use the definition \eqref{defR} to write the separated correlator in terms of nested commutators. As discussed in section \ref{sec:DifferentORderings}, most of the terms drop out using the fact that the ANE annihilates the vacuum. All that remains is a Wightman function and we obtain 
\begin{align}\label{awithpc}
\int_{v_1<0} &d^4x_1 \int_{v_2<0} d^4 x_2\,  e^{ik_1 \cdot x_1 + i k_2 \cdot x_2}\langle \Theta(x_1) \E_u(0) \Theta(x_2)\rangle + PC \\
&\approx 
8\pi k_{1u}^2\Big[ (a_{IR} - a_{UV})(k_1+k_2)^2  - \tfrac{3}{2} b_1(k_1^2+k_2^2) - \tfrac{3}{2}b_2 M^2 \Big] \delta(k_{1u} + k_{2u})
\notag \ .
\end{align}
The `sep' subscript has been dropped because the Wightman function does not have contact terms; there are no coincident point contributions to the first term.

Our goal now is to relate $\Delta a = a_{UV} - a_{IR}$ to the separated 3-point function. This is complicated by the presence of the unknown partial contact term $PC$, which has no known positivity properties because it involves coincident points. Below in section \ref{ss:PC} we will analyze the partial contact terms and show that they arise only when there is a conserved dimension-4, spin-2 operator other than the stress tensor, i.e., a possible improvement term. Typically the stress tensor is the unique primary operator with these quantum numbers, but there can also be contributions from $(\p_\alpha \p_\beta-g_{\alpha\beta}\p^2) \O$ where $\O$ is a scalar primary of dimension two. The contribution of such an operator --- and therefore the most general allowed form of the partial contact term --- is
\begin{align}\label{PCF}
PC = [F(k_1) + F(k_2)] k_{1u}^2 \delta(k_{1u} + k_{2u})\, ,
\end{align}
with $F$ an unknown function. Crucially, this has no $k_1 \cdot k_2$ momentum dependence. We can therefore extract $\Delta a$ from \eqref{awithpc} by picking off the coefficient of $k_1 \cdot k_2$. 

The Fourier transform in \eqref{awithpc} may be divergent, and then the split into separated and contact terms becomes ambiguous. However the coefficient of the $O(k^4)$ term converges and is therefore  defined unambiguously.

\subsubsection{The $a$-theorem}

We seek a positive sum rule using the ANEC in the form
\begin{align}
\int_{v_1<0} d^4 x_1 \int_{v_2<0} d^4 x_2 \, f^*(x_1) f(x_2)\langle \Theta(x_1)\E_u(0)\Theta(x_2)\rangle \geq 0 \ .
\end{align}
This inequality holds for any smearing kernel $f$. The simplest choice is 
\begin{align}\label{fkernel}
f_m(x) = e^{-i \omega u} x^{\perp}_m\, ,
\end{align}
where $m=1,2$ runs over the transverse directions, i.e. $\vec{x} = (x^\perp_1, x^\perp_2)$.  This kernel is equivalent to acting with a derivative $(-\p_{\vec{k}_1} \cdot \p_{\vec{k}_2})$ on \eqref{awithpc}, then setting $k_{1u} = - k_{2u} = \omega$ and $k_{1v}=k_{2v} = \vec{k}_1 = \vec{k}_2 = 0$. The leading term at small $\omega$ is therefore
\begin{align}\label{srWithV}
\int_{v_1<0}d^4 x_1 \int_{v_2<0} d^4 x_2\, e^{i\omega(u_1-u_2)}\vec{x}_1 \cdot \vec{x}_2 \langle\Theta(x_1)\E_u(0)\Theta(x_2)\rangle \approx 32 \omega^2 (a_{UV} - a_{IR}) V^u\, ,
\end{align}
where $V^u = \pi \delta(0) = \frac{1}{2}\int du$ is an infinite volume factor. The $\vec{k}$-derivatives have removed any contributions from the non-universal terms $PC$, $b_1$, and $b_2$.

The sum rule \eqref{srWithV} relates $\Delta a$ to an expectation value of $\E_u(0)$, which is required to be non-negative by the ANEC. Therefore, the ANEC implies the $a$-theorem:
\begin{align}
a_{UV} \geq a_{IR} \ .
\end{align}
The $a$-theorem was first derived by Komargodski and Schwimmer by matching the anomaly in the dilaton scattering amplitude, which is related the four-point function of the trace \cite{Komargodski:2011vj}. This followed a study of anomaly matching across phases with spontaneously broken conformal symmetry in \cite{Schwimmer:2010za}. The strategy we have followed here is also similar to anomaly matching, but applied to the stress tensor 3-point function rather than the 4-point function. 

There are two small variants of the sum rule that we will discuss in the following subsections. The first is an IR-regulated version using a wavepacket, which replaces $V^u$ in the sum rule by a finite positive number --- this justifies ignoring the infinite volume factor in the derivation of the $a$-theorem. The second is another way of regulating the volume factor by fixing the $u$-position of one of the operator insertions. After describing these other sum rules we will return to the analysis of partial contact terms in section \ref{ss:PC} in order to justify the claim \eqref{PCF}.

\subsubsection{Wavepacket regulator}
Define the state
\begin{align}\label{psistate}
|\psi(\omega)_m\rangle &= \int d^4 x\, \theta(-v) e^{i \omega u - u^2/\sigma^2}x^\perp_m\, \Theta(u,v,\vec{x})|0\rangle \ , 
\end{align}
where $\sigma$ is an infrared cutoff with $\sigma^{-1} \ll \omega \ll M$. This is designed to be an IR-regulated version of the smearing in \eqref{fkernel}. Using \eqref{awithpc} we find
\begin{align}\label{positivesumrule4d}
\sum_{m=1,2}\langle \psi(\omega)_m | \E_u(0) | \psi(\omega)_m\rangle &= 
\int_{v_1<0} d^4 x_1 \int_{v_2<0} d^4 x_2\, e^{i\omega (u_2-u_1) - (u_1^2+u_2^2)/\sigma^2} \vec{x}_1 \cdot \vec{x}_2 \langle \Theta(x_1) \E_u(0) \Theta(x_2)\rangle\notag\\
&\approx 8 (a_{UV} - a_{IR}) \sqrt{2\pi} \sigma \omega^2 \ .
\end{align}
To obtain the second line, the Gaussian damping factors have been written as  $e^{-u^2/\sigma^2} = \frac{\sigma}{2\sqrt{\pi}} \int dp\, e^{ipu - p^2 \sigma^2/4}$ before exchanging the orders of integration. Comparing to \eqref{srWithV}, we see that the volume factor $V^u$ has been replaced by a finite positive number as expected.

\subsection{Exact sum rules for $\Delta a$}
The volume divergence $V^u$ can also be regulated by fixing the $u$-position of one of the operator insertions and using null translation invariance. In addition, we can extract the $\omega^2$ contribution to the correlator by acting with $\p_\omega^2$ and setting all momenta to zero. If we fix $u_2 =0 $ in \eqref{srWithV}, this gives the exact sum rule
\begin{align}\label{exactR}
a_{UV} - a_{IR} &=  - \frac{1}{32} \int_{v_1<0} d^4 x_1 \int_{v_2<0} d^4 x_2 \, u_1^2 \delta(u_2) \vec{x}_1 \cdot \vec{x}_2 \langle \Theta(x_1)\E_u(0)\Theta(x_2)\rangle \, .
\end{align}
Once again there are no contact terms in the integral.
This formula can also be derived directly by starting with the CFT equation \eqref{aCFT}, as applied to the IR theory, and following the same logic as section \ref{ss:deriveAsum} --- that is, moving the UV contact terms to the left-hand side.

A second version of the exact sum rule comes from applying the same logic to the time-ordered correlator. The contact term just differs by a sign, so the result is
\begin{align}\label{exactT}
a_{UV} - a_{IR} &=   \frac{1}{32} \int_{v_1>0} d^4 x_1 \int_{v_2<0} d^4 x_2 \, u_1^2 \delta(u_2) \vec{x}_1 \cdot \vec{x}_2 \langle \Theta(x_1)\E_u(0)\Theta(x_2)\rangle \, .
\end{align}

Yet another option is to fix the $u$-position of $T_{uu}$ rather than the trace, as in \eqref{aCFTlocal}. This gives the symmetrical sum rule
\begin{align}\label{exactLocal}
a_{UV} - a_{IR} &= -\frac{1}{32}\int_{v_1<0} d^4 x_1 \int_{v_2<0} d^4 x_2 (u_1-u_2)^2 \vec{x}_1 \cdot \vec{x}_2 \langle \Theta(x_1) T_{uu}(0) \Theta(x_2)\rangle \ .
\end{align}

None of the exact sum rules \eqref{exactR}-\eqref{exactLocal} is manifestly positive, because they are not expressed as an expectation value of $\E_u$. Nonetheless they are positive by the results of section \ref{ss:deriveAsum}.

\subsection{Analysis of partial contact terms}\label{ss:PC}

It remains to justify dropping partial contact terms in the derivation of the sum rule for $\Delta a$.

 In $d$ dimensions, consider the Euclidean correlation function
\begin{align}
\langle T_{\alpha\beta}(x_3) \Theta(x_1)\Theta(x_2)\rangle \ . 
\end{align}
Partial contact terms, by definition, are contributions with two points coincident, and the third point separated. There are two different types of partial contact terms to consider: terms with $(\Theta\Theta)$ coincident, involving $\delta^{(d)}(x_1-x_2)$, and terms with $(T_{\alpha\beta}\Theta)$ coincident, involving $\delta^{(d)}(x_3-x_1)$ or $\delta^{(d)}(x_3-x_2)$. Both types of partial contact terms can exist (and they are nonzero in the free massive scalar), but we will show that they do not contribute to the sum rule. 

Let us start with the first type, where the two traces $\Theta$ coincide. It is easy to see that any such terms drop out of the ANE correlators $\langle \R[\E_u(0); \Theta(x_1)\Theta(x_2)]\rangle$ or $\langle \T[\E_u(0) \Theta(x_1)\Theta(x_2)]\rangle$. In the case where the two traces $\Theta$ coincide, they must be ordered on the same side of the ANE insertion. Therefore we can use $\E_u|0\rangle = 0$ and these contributions vanish.

We now turn to the partial contact terms where $T_{\alpha\beta}$ coincides with one of the $\Theta$ insertions.
Such a term arises if there is a delta function in the OPE,
\begin{align}
T_{\alpha\beta}(x_3)\Theta(x_1) \supset \O^{\sigma_1\dots\sigma_\ell}
(x_1)D_{\alpha\beta\sigma_1\dots\sigma_\ell}(\p)\delta^{(d)}(x_3-x_1)\, , 
\end{align}
where $\O$ is a (not necessarily primary) local operator in the UV CFT (multiplied by $\beta$-functions), $D$ is a differential operator, and $\ell \geq 0$. 
Therefore in the 3-point function they take the form
\begin{align}
\langle T_{\alpha\beta}(x_3) \Theta(x_1)\Theta(x_2)\rangle
&\supset  \langle \O^{\sigma_1\sigma_2\dots \sigma_\ell}(x_1) \Theta(x_2)\rangle D_{\alpha\beta \sigma_1\sigma_2\dots \sigma_\ell}(\p) \delta^{(d)}(x_3-x_1) \\
&\qquad  + (1\leftrightarrow 2)\notag\, .
\end{align}
In the Lorentzian 3-point function in momentum space, focusing on the time-ordered correlator for concreteness (the retarded correlator is similar), this becomes
\begin{align}\label{tpc1}
\lll \T[ T_{\alpha\beta}(k_3) \Theta(k_1) \Theta(k_2)]\rr
&\supset G_{\O\Theta}^{\sigma_1\dots \sigma_\ell}(k_2) P_{\alpha\beta\sigma_1\dots\sigma_\ell}(k_3) + (1\leftrightarrow 2) \, .
\end{align}
Here $P$ is a tensor that is an analytic function of $k_3$, and
\begin{align}
G_{\O\Theta}^{\sigma_1\dots \sigma_\ell}(k) &= \lll \T[\O^{\sigma_1\dots\sigma_\ell}(k)\Theta(-k)]\rr \, .
\end{align}
The correlator in the sum rule has $k_{3u} = 0$ and only involves the null energy, $T_{uu}$, so we now specialize to this case, and denote
\begin{align}
\Gamma &\equiv \left.\lll \T[ T_{uu}(k_3) \Theta(k_1) \Theta(k_2)]\rr_{PC} \right|_{k_{3u}=0}\, .
\end{align}
The tensor indices on $P$ must be accounted for by combinations of $k_{3}$'s and metric tensors. If $\ell = 0$ or 1 and $k_{3u}=0$, it is impossible to write a nonzero tensor structure for $P_{uu\sigma_1\dots\sigma_\ell}$. For $\ell \geq 2$, the nonzero part of this tensor must take the form
\begin{align}
\left.P_{uu\sigma_1 \sigma_2 \dots \sigma_{\ell}}(k_3)\right|_{k_{3u}=0}
= g_{u\sigma_1}g_{u\sigma_2} U_{\sigma_3\dots\sigma_\ell}(k_3) \, ,
\end{align}
for some tensor $U$.
To see this, we note that each $u$ index must appear on a metric tensor because $k_{3u}=0$, but they cannot appear together, because $g_{uu}=0$. 
Thus we have shown that any partial contact term must enter the 3-point function in the form
\begin{align}\label{tpc2}
\Gamma
&\sim 
\lll \T[ \O_{uu}{}^{\sigma_3\dots \sigma_\ell}(k_2)\Theta(-k_2)]\rr
U_{\sigma_3\dots\sigma_\ell}(k_3)
+ (1\leftrightarrow 2) \ .
\end{align}
The sum rule for $\Delta a$ was designed to pick off the term proportional to $k_{2u}^2 k_1 \cdot k_2$.
The 2-point function on the right-hand side of \eqref{tpc2} provides the factor of $k_{2u}^2$. There are only two ways to get a factor of $k_1 \cdot k_2$, which are the following terms with $\ell=2$ and $\ell=3$:
\begin{align}
\Gamma &\sim 
\lll \T[\O_{uu}(k_2)\Theta(-k_2)\rr k_3^2 + \lll \T[\O_{uu}{}^\sigma(-k_2)\Theta(k_2)\rr k_{3\sigma} + (1\leftrightarrow 2) \, .
\end{align}
By dimensional analysis, the first term only affects the sum rule if the operator has scaling dimension $\Delta(\O_{uu}) = 2$ and the second term only affects the sum rule for $\Delta(\O_{uu}{}^\sigma) = 3$. However, this is forbidden by the unitarity bound in the UV CFT. Therefore there are no partial contact contributions to the sum rule. This also justifies \textit{a posteriori} using the split \eqref{tsplit} in the derivation.

One may wonder whether any partial contact terms at all are allowed in the correlation function $\lll \T[T_{uu}(k_3)\Theta(k_1)\Theta(k_2)]\rr$ when $k_{3u} =0$. It follows from \eqref{tpc2} and the unitarity bounds that the only partial contact term at $O(k^4)$ has $\ell=2$,  $U = \mbox{constant}$, and $\Delta(\O_{uu}) = 4$. This can arise if there is a dimension-2 scalar field, with $\O_{\mu\nu} = (\p_\mu\p_\nu - g_{\mu\nu}\p^2)\O$. Because of the momentum dependence, this does not affect the sum rule for $\Delta a$, but it prevents us from a writing a sum rule for the non-universal terms  $\Delta b_1$ and  $\Delta b_2$ in terms of the separated correlator.

\section{Example: Massive scalar\label{sec:MassiveScalar}}
In this section, we will apply the sum rule to a conformally coupled free massive scalar field in four dimensions. This theory flows to a massless scalar in the UV, with $a_{UV} = \frac{1}{5760\pi^2}$ (see e.g. \cite{Komargodski:2011vj}). In the IR it flows to the trivial empty CFT with $a_{IR}=0$.
%
The action in $d$ dimensions is 
\begin{align}
S = - \frac{1}{2}\int d^d x \sqrt{-g} ( (\p \phi)^2 + m^2 \phi^2 + \xi R \phi^2)\, ,
\end{align}
where the conformal coupling is $\xi = \frac{d-2}{4(d-1)}$. The stress tensor in flat spacetime derived from this action is
\begin{align}
T_{\mu\nu} &= \p_\mu \phi \p_\nu\phi - \frac{1}{2} g_{\mu\nu}(m^2\phi^2 +(\p \phi)^2)  
-\xi (\p_\mu \p_\nu - g_{\mu\nu}\p^2)\phi^2\, ,\label{eq:TarbitraryD}
\end{align}
and the trace is 
\be 
\Theta = -\frac{d}{2}m^2 \phi^2 +2(d-1) \xi \phi \Box \phi + 
\left( 2(d-1)\xi - \frac{d}{2}+1\right) (\p \phi)^2 \ .\label{eq:TracearbitraryD}
\ee 
We will compute time-ordered connected correlation functions involving the stress tensor $T_{\mu\nu}$ and its trace. At separated points these are given by Wick contractions, using the Feynman propagator 
\begin{align}
G(x-y) = \langle{\cal T}[ \phi(x)\phi(y)] \rangle = -i \int \frac{d^dp}{(2\pi)^d} \frac{e^{ip\cdot(x-y)}}{p^2+m^2 - i\epsilon}\, .
\end{align}
When $d=4$, the conformal coupling is $\xi = \frac{1}{6}$ such that the trace is
\begin{align}
\Theta &= -m^2 \phi^2 + \phi(\Box-m^2)\phi \, ,\label{eq:traceMS}
\end{align}
and the null energy is
\begin{align}
T_{uu} &= \frac{2}{3} (\p_u \phi)^2 - \frac{1}{3} \phi \p_u^2 \phi \ .
\end{align}


In the calculation of correlation functions at separated points, we are free to use the equation of motion to replace the trace by
\begin{align}
\Theta\quad  \to  & \quad -m^2 \phi^2 \ ,\label{eq:NEwTrace}
\end{align}
because terms like $\langle \phi(x) (\Box-m^2) \phi(y)\rangle$ can only produce contact terms.
We can then use Wick contractions (with the substitution \eqref{eq:NEwTrace}) to obtain the correlator at separated points. This gives
\begin{align}\label{tttsep}
\langle {\cal T}[ T_{uu}(x_3)\Theta(x_1)\Theta(x_2)]\rangle_{\rm sep} &= 
\frac{4m^4}{3} G(x_{12})\left( 4\p_{u_1} \p_{u_2} - \p_{u_1}^2 - \p_{u_2}^2\right)\left[ G(x_{13}) G(x_{23}) \right]\, ,
\end{align}
where $x_{ij}=x_i-x_j$, which is a product of Bessel functions.

\begin{figure}[t]
\begin{center}
\begin{overpic}[grid=no,scale=0.5]{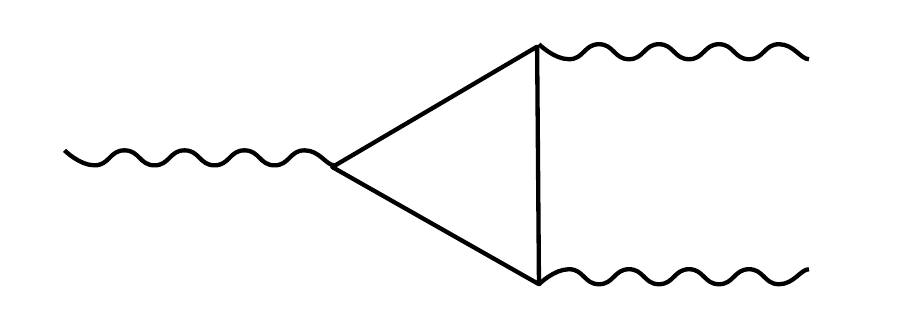}
\put (-100,180) {$T_{uu}$}
\put (950, 280) {$\Theta$}
\put (950,40) {$\Theta$}
\end{overpic}
\end{center}
\caption{\small Feynman diagram for $\langle {\cal T}[ T_{uu}(k_3) \Theta(k_1) \Theta(k_2) ]\rangle$ in the free massive scalar.\label{fig:TuuThetaTheta}}
\end{figure}

We will check the time-ordered version of the sum rule in \eqref{exactT}. The retarded sum rule is the same up to a minus sign. To evaluate them, we will first go to momentum space.
The Fourier transform of \eqref{tttsep} is
\begin{align}\label{tuuscalarDiagram}
\lll {\cal T}[T_{uu}(k_3)&\Theta(k_1)\Theta(k_2)] \rr_{\rm sep}  \\
&=
\frac{4m^4}{3} i \int \frac{d^4p}{(2\pi)^4} \frac{ (p_u+ k_{1u})^2 + (p_u - k_{2u})^2 + 4  (p_u + k_{1u})(p_u - k_{2u})}{ (p^2 + m^2 - i \epsilon)
( (p+k_1)^2+m^2 - i\epsilon)( (p-k_2)^2 + m^2 - i \epsilon )} \notag \, .
\end{align}
This is the 1-loop diagram shown in figure \ref{fig:TuuThetaTheta}, 
up to the contact terms that we dropped by applying the equations of motion. 
The result of the integral at low external momentum is
\begin{align}\label{tuuscalar}
\lll {\cal T}[T_{uu}(k_3)&\Theta(k_1)\Theta(k_2)] \rr_{\rm sep}  \\
&= - \frac{m^2}{24\pi^2} k_{1u} k_{2u} + \frac{1}{720\pi^2}(k_1+k_2)^2 (k_{1u}+k_{2u})^2 \notag \\
&\quad -8 \cdot \frac{1}{5760\pi^2}(k_{1u}^2k_2^2+k_{2u}^2 k_1^2 - 2 k_{1u} k_{2u} k_1\cdot k_2) \notag \\
&\quad  - 4 \cdot \frac{1}{2880\pi^2}( (k_{1u}+k_{2u})^2k_1\sdot k_2  - 3 k_{1u} k_{2u}(k_1^2+k_2^2))      + O(k^6) \ ,\notag
\end{align}
where we have expanded in $\frac{k_1^2}{m^2}, \,\frac{k_2^2}{m^2}, \,\frac{k_1\cdot k_2}{m^2} \ll 1$.
 The integral is done by Wick rotating $p^t \to - i p^\tau$, combining the denominators with the Feynman trick, expanding in $k$, then doing the Feynman parameter integrals and lastly the convergent loop integral.\footnote{
Initially the loop diagram appears to be log divergent, but the terms with $p_u^2$ in the numerator drop out because they have integral proportional to $g_{uu} = 0$. In any case, only the $k$-derivatives of the correlator appear in the sum rule, and after taking a $k$-derivative the loop integral converges. Note that if we had not used the equations of motion, the integral would have power-law divergences and we would need to renormalize to calculate the full correlator. Thus the fact that the sum rule only involves the correlator at separated points is a major simplification in this calculation. Note that because of the UV divergences, the split into separated and contact terms in the $k^2m^2$ term is ambiguous, but the calculation of the separated $k^4$ terms is well defined. 
}
Although regularization is not necessary, it can be a convenient way to check the integral, and in appendix \ref{app:loops} we reproduce this result (and a few other similar correlators in the massive scalar theory) using dimensional regularization.

Finally we evaluate the sum rule \eqref{exactT}, 
\begin{align}
\Delta a &= \frac{1}{64} \int du_3 \int d^4 x_1 d^4 x_2\, u_1^2 \delta(u_2) \vec{x}_1 \sdot \vec{x}_2 \langle {\cal T}[ T_{uu}(u_3,v_3=0,\vec{x}_3=0) \Theta(x_1) \Theta(x_2)] \rangle_{\rm sep} \\
&=  \frac{1}{64}  \left. (\p_{k_{1u}}-\p_{k_{2u}})^2 \vec{\p}_{k_1} \cdot \vec{\p}_{k_2} \lll  {\cal T}[T_{uu}(-k_1-k_2)\Theta(k_1)\Theta(k_2)] \rr_{\rm sep}  \right|_{k_1=k_2=0}
\notag\\
&= \frac{1}{5760\pi^2}\, ,
\end{align}
which is indeed equal to the Euler anomaly $a_{UV}$ for a free massless scalar.

\section{Discussion}\label{s:discussion}

\subsection{Technical summary}

In this subsection we review the derivation of the sum rule and the $a$-theorem to emphasize the main steps. We will ignore a variety of non-universal contributions to the correlation functions that appear in the intermediate steps. All of the non-universal terms were considered in detail above and shown to drop out of the final sum rule.


We studied the retarded correlation function
\begin{align}\label{introret}
\langle \R[\E_u; \Theta(x_1)\Theta(x_2)]\rangle &:= -\theta(-v_1)\theta(v_1-v_2)\langle [[\E_u,\Theta(x_1)],\Theta(x_2)]\rangle  - (x_1 \leftrightarrow x_2) + \mbox{contact terms} \, ,
\end{align}
where $\Theta = T\indices{_\mu^\mu}$ and $\E_u = \int du T_{uu}(u,v=0, \vec{x}=0)$. 
The starting point is the conformal anomaly, $\langle \Theta\rangle = c (\mbox{Weyl})^2 -  a(\mbox{Euler})$. By varying the anomaly with respect to the background metric, we derived the contact terms in the $\langle T_{\alpha\beta}\Theta\Theta\rangle$ correlation function. With the trace insertions in momentum space, this led to 
\begin{align}\label{conclusionRcft}
\langle \R[ \E_u; \Theta(k_1)\Theta(k_2) ]\rangle_{\rm CFT} = 16 \pi a k_{1u}^2(k_1+k_2)^2\delta(k_{1u} + k_{2u}) + \cdots
\end{align}
where the dots are non-universal. See \eqref{answerE4d} for the complete formula. 

The result \eqref{conclusionRcft} holds at a conformal fixed point. In a quantum field theory that flows between two fixed points, the correlators of the QFT at small momentum agree with the infrared CFT. Therefore \eqref{conclusionRcft} holds for the QFT correlator at small momentum, with $a_{IR}$ appearing on the right-hand side:
\begin{align}\label{conclusionRIR}
\langle \R[ \E_u; \Theta(k_1)\Theta(k_2) ]\rangle_{\rm QFT} \approx 16 \pi a_{IR} \, k_{1u}^2(k_1+k_2)^2\delta(k_{1u} + k_{2u}) + \cdots
\end{align}
(There are other, non-universal terms at the same order in $k$ that drop out of the sum rule and were discussed above.) 
The left-hand side has a contribution from the UV region of the Fourier integral, which is controlled by the UV fixed point. Moving the UV contact terms to the other side of the equation we obtain
\begin{align}\label{conclusionRdiff}
\langle \R[ \E_u; \Theta(k_1)\Theta(k_2) ]\rangle_{\sep} = 16 \pi (a_{IR} - a_{UV}) (k_1+k_2)^2 \delta(k_{1u} + k_{2u}) + \cdots \ .
\end{align}
with `sep' for `separated' indicating that no contact terms are included in the Fourier integral that defines the left-hand side. Thus the left-hand side, being a retarded correlation function at separated points, is the Fourier transform of the nested commutator
\begin{align}
-\theta(-v_1)\theta(v_1-v_2) \langle [[\E_u(0), \Theta(x_1)], \Theta(x_2)]\rangle + (1\leftrightarrow 2)
\end{align}
The ANE annihilates the vacuum, $\E_u|0\rangle =0$, so when the commutators are expanded, most of the terms drop out. What survives is a Wightman function,
\begin{align}\label{asumi2}
&\int_{v_1<0}d^4x_1\,\int_{v_2<0} d^4 x_2\, e^{ik_1 \cdot x_1 + i k_2 \cdot x_2} \langle \Theta(x_1)\E_u(0)\Theta(x_2)\rangle
\\
&\qquad \qquad \qquad\qquad = -8\pi k_{1u}^2  (a_{UV} - a_{IR})(k_1+k_2)^2\delta(k_{1u}+k_{2u}) + \cdots \ .\notag
\end{align}
For the complete version of this formula including non-universal terms see \eqref{awithpc}. 
To prove positivity, act on both sides with $ -\vec{\p}_{k_1} \cdot \vec{\p}_{k_2}$ and set $k_{2u} = -k_{1u}$. The left-hand side becomes an expectation value $\langle \Psi | \E_u |\Psi\rangle$, which is positive by the ANEC, and the right-hand side is proportional to $a_{UV} - a_{IR}$. The non-universal terms in `$\dots$' drop out of the leading term at low momentum because they do not have $\vec{k}_1 \cdot \vec{k}_2$ dependence. 
It follows that $a_{UV} \geq a_{IR}$.   The sum rule \eqref{introsum4} is the coefficient of $k_{1u}^2 \vec{k}_1 \cdot \vec{k}_2$ in \eqref{asumi2}. Three slightly different versions of the sum rule are discussed above: A manifestly positive, IR-regulated expression for $\Delta a$ in \eqref{positivesumrule4d}, a time-ordered sum rule involving the averaged null energy in \eqref{exactT} and a retarded sum rule in \eqref{exactR}.

In this summary we have glossed over two crucial details that were discussed at length above. The first is the analysis of  counterterms, which is necessary to show that the non-universal terms drop out. The second is the issue of `partial contact' terms, which are contributions to the QFT 3-point function with two points coincident and one point separated. We have shown that all of the non-universal terms and partial contact terms drop out of the sum rule.

\subsection{Connections to quantum information}
Using the results of \cite{Faulkner:2016mzt} relating the ANE to the modular Hamiltonian, the derivation of the $c$-theorem and $a$-theorem from the ANEC can be re-phrased in terms of the relative entropy: Monotonicity of relative entropy implies monotonicity of the renormalization group in two and four dimensions. This is distinct from the information-theoretic derivation of the $C$-theorems in \cite{Casini:2004bw,Casini:2006es,Casini:2017vbe,Casini:2023kyj}. 

To make this more explicit, let us review the derivation of the ANEC in \cite{Faulkner:2016mzt}. In null coordinates $ds^2 = -dudv + d\vec{x}^2$, let region $A$ be the Rindler wedge $u>0$, $v<0$, and $A^c$ the complementary Rindler wedge $u<0$, $v>0$. The full modular Hamiltonian is defined
\begin{align}
\widehat{H}_A = H_A - H_{A^c}
\end{align}
where $H_A$ is the modular Hamiltonian for region $A$ in the vacuum state. The density matrix of Rindler space is thermal with respect to the boost generator, so $\widehat{H}_A$ is an integral of the stress tensor. 

The two regions $A$ and $A^c$ meet at the Rindler bifurcation surface, $u=v=0$. We now consider deforming this surface in the null direction, so that the regions end at $v=0$, $u = \epsilon(\vec{x})$. Under this deformation the  change in the modular Hamiltonian is \cite{Faulkner:2016mzt} 
\begin{align}\label{defmod}
\delta \widehat{H}_A = - 2\pi \int d^{d-2}\vec{x} \epsilon(\vec{x}) \int du T_{uu}(u, v=0, \vec{x}) \ .
\end{align}
Consider the relative entropy $S(\rho_A^\psi||\rho_A)$ where $\rho_A$ is the vacuum state and $\rho_A^{\psi}$ is an excited state. This quantity is monotonic under partial trace, which in the present context means a deformation of the Rindler wedge with $\epsilon(\vec{x})>0$. Relative entropy is a measure of distinguishability, and monotonicity captures the intuition that hiding part of a system can only make it harder to distinguish two states. Monotonicity of relative entropy can be used to prove that the full modular Hamiltonian satisfies $\delta \langle \psi|\widehat{H}_A |\psi\rangle < 0$ under such a deformation. Together with \eqref{defmod}, this implies the ANEC \cite{Faulkner:2016mzt}.

In \eqref{psistate}, we defined a state $|\Psi\rangle$ created by a smeared insertion of the trace $\Theta$ such that $\langle \Psi | \E_u |\Psi\rangle$ is proportional to $a_{UV} - a_{IR}$. Therefore in this state the monotonicity of relative entropy under a null deformation of the half-space implies the $a$-theorem.

Relative entropy was used in a different way in \cite{Casini:2016udt} to prove the 2d $c$-theorem and, in higher dimensions, to constrain the RG flow of the area term in the entropy. Distinguishability and relative entropy have also been discussed as a measure of distance between quantum field theories in \cite{Balasubramanian:2014bfa,Stout:2021ubb,Erdmenger:2021sot,Stout:2022phm}.

\subsection{Connections to Lorentzian inversion?}

The averaged null energy condition was derived by CFT methods in \cite{Hartman:2016lgu} (see also \cite{Hartman:2015lfa,Hartman:2016dxc,Kravchuk:2018htv}). Those methods are now best understood as a particular limit of the Lorentzian inversion formula \cite{Caron-Huot:2017vep,Simmons-Duffin:2017nub}. The relation is that the positive sum rule for the ANE derived in \cite{Hartman:2016lgu} is the leading-twist term in the inversion formula.

This suggests a role for Lorentzian inversion in the study of RG flows. Of course, the formula of Caron-Huot applies only to conformal field theories. However, if it is applied to the infrared fixed point, it has contributions from all scales, and therefore encodes information about the RG flow from the UV. Perhaps this can be related to the dispersion relation for the dilaton scattering amplitude used to derive the $a$-theorem in \cite{Komargodski:2011vj}. If so, then the inversion formula, which relates 3-point functions to 4-point functions, may provide a bridge between our approach and the derivation of Komargodski and Schwimmer \cite{Komargodski:2011vj}. 
 It would also be very interesting to connect to the local renormalization group \cite{Osborn:1989td,Jack:1990eb,Jack:2013sha}, which  has been applied to this problem in order to relate the $2 \to 2$ dilaton $S$-matrix to 3-point amplitudes \cite{Baume:2014rla,Shore:2016xor}.

To expand on this, let us compare our sum rule to that of \cite{Komargodski:2011vj}. By simply setting the two results for $\Delta a$ equal, we obtain the suggestive relation
\begin{align}\label{bothsumrules}
- \frac{1}{32} \int_{v_1<0} d^4 x_1 \int_{v_2<0} d^4 x_2 \, u_1^2 \delta(u_2) \vec{x}_1 \cdot \vec{x}_2 \langle \Theta(x_1)\E_u(0)\Theta(x_2)\rangle 
&= \frac{f^4}{\pi} \int_0^\infty ds \frac{ \mbox{Im}\, {\cal A}(s) }{s^3} \ , 
\end{align}
where ${\cal A}$ is the dilaton scattering amplitude in the forward limit. This is interesting because it hints at a generalized Lorentzian inversion formula for RG flows.

To see why, let us briefly review the relationship between light-ray operators and the inversion formula.  Following \cite{Hartman:2016lgu}, consider a 4-point function of identical scalar primaries of dimension $\Delta$,
\begin{align}
G = \langle\O(u,v) \O(x_1)   \O(x_2)  \O(-u,-v) \rangle \ ,
\end{align}
where we have set the transverse coordinates to zero in the first and last insertions. Define
\begin{align}
\eta = -u v , \qquad \sigma = 1/u
\end{align}
and choose kinematics with $0 < \eta \ll \sigma \ll 1$. 
We also assume $x_1$ is in the left Rindler wedge, and $x_2$ is in the right Rindler wedge. In the  limit $\eta \to 0$ at fixed $u$, two of the insertions become null separated. 
Using the lightcone OPE, the following sum rule for the averaged null energy was derived in \cite{Hartman:2016lgu}:\footnote{This equation and its higher-spin generalization can be found in eqn (6.4) in \cite{Hartman:2016lgu}. The normalization factor $N$ is a constant that can be found there. We have used the relation Re disc $G(z,\bz) = \mbox{dDisc} G(z,\bz)$ for $z,\bz \in (0,1)$ to rewrite the integrand it in terms of the double discontinuity.
}
\begin{align}\label{lcsumrule}
\langle \O(x_1) \E_u(0) \O(x_2)\rangle
&= N \lim_{\epsilon\to 0} \lim_{\eta \to 0} \eta^{\Delta - (d-2)/2}
\int_{-\epsilon}^{\epsilon} d\sigma \,\mbox{dDisc} \, G 
\end{align}
The integrand is the double discontinuity, which is a non-negative double commutator \cite{Maldacena:2015waa,Hartman:2015lfa},
\begin{align}
\mbox{dDisc} \, G &= \langle [\O(x_1), \O(u,v)][\O(-u,-v),\O(x_2)]\rangle \geq 0 \ .
\end{align}
The light-ray sum rule \eqref{lcsumrule} is a particular limit of the Lorentzian inversion formula. To see this, one simply expands the integrand in the inversion formula in the same kinematic limit, and this reproduces exactly \eqref{lcsumrule} \cite{Kravchuk:2018htv,tomBootstrapLectures}.

Now let us compare the CFT sum rule \eqref{lcsumrule} to the RG sum rule \eqref{bothsumrules}. In both cases, we have a light-ray expectation value on the left, and a manifestly-positive quantity built from the 4-point function on the right. This is the sense in which we view the RG sum rule as a natural extension of the lightcone inversion formula to an RG flow. It would be interesting to understand this more directly, and perhaps to find a more general inversion formula beyond CFT.

\ \\

\noindent \textbf{Acknowledgments}

\noindent We are grateful to Simon Caron-Huot, Horacio Casini, Jeevan Chandra, Clay Cordova, Diego Hofman, Austin Joyce, Denis Karateev, Murat Kologlu, Zohar Komargodski, Juan Maldacena, David Meltzer, Ian Moult, Joao Penedones, Shu-Heng Shao, Nathan Seiberg, David Simmons-Duffin, and John Stout for helpful discussions. This work is funded by NSF grant PHY-2014071. We also acknowledge support from NSF grant PHY-1748958 for participation in a KITP workshop. Part of this work was performed in part at Aspen Center for Physics, which is supported by NSF grant PHY-2210452.

\addtocontents{toc}{\protect\setcounter{tocdepth}{1}}

\appendix

\section{Ward identities}\label{ap:WardIdentities}
In this appendix we derive the Ward identities and explain some subtleties associated to the transformation of the stress tensor under diffeomorphisms. Everything in this appendix is in Euclidean signature.

\subsection{Diffeomorphisms}
The subtleties stem from the fact that with the standard Osborn-Petkou conventions \eqref{eq:Tcorrelator1}, used throughout the paper, the stress tensor does not quite behave like a tensor. Although $\langle T_{\mu\nu}\rangle$ is an ordinary (weight zero) tensor, the operator $T_{\mu\nu}$ inside a correlation function behaves instead like a tensor density of weight one. The reason is that correlation functions are defined with the $1/\sqrt{g}$'s  multiplied at the end, rather than inside the variations. Thus  $T_{\alpha\beta}$ defined following Osborn and Petkou is actually a tensor density that is multiplied by an inert, non-transforming factor of $1/\sqrt{g}$ at the end of a calculation.\footnote{ It is possible to define a stress tensor that behaves like a true tensor inside correlation functions by bringing the $1/\sqrt{g}$'s inside the variations, but at the expense of producing asymmetric contact terms. This is why we chose to use the Obsorn-Petkou tensor density conventions. The resulting Ward identities \cite{Osborn:1993cr} look superficially different from those found in many other sources including the textbooks e.g.~\cite{DiFrancesco:1997nk}  which use tensor conventions. }

To derive the Ward identities for the stress tensor it is convenient to start with the stress tensor density,
\begin{align}
\tT_{\mu\nu} = \sqrt{g} T_{\mu\nu} \ ,
\end{align}
with correlation functions
\begin{align}
\langle \tT_{\mu\nu}(x_1) \tT_{\alpha\beta}(x_2) \cdots\rangle
&= (-2)^n \frac{\delta}{\delta g^{\mu\nu}(x_1)} \frac{\delta}{\delta g^{\alpha\beta}(x_2)} \cdots \log Z\, .
\end{align}
$\tT_{\mu\nu}$ is a tensor density of weight one.
In general, under a diffeomorphism $x^\mu \to x^\mu - \zeta^\mu$, the transformation of a tensor density $\O$ of weight $w$ is the Lie derivative
\begin{align}
{\cal L}_\zeta \O^{\alpha_1\alpha_2\dots}{}_{\beta_1\beta_2\dots}
&= \zeta^\mu \p_\mu \O^{\alpha_1\alpha_2\dots}{}_{\beta_1\beta_2\dots}
- \p_\mu \zeta^{\alpha_1} \O^{\mu \alpha_2\dots}{}_{\beta_1\beta_2\dots}
-\cdots\\
&\quad
+ \p_{\beta_1} \zeta^\mu \O^{\alpha_1\alpha_2\dots}{}_{\mu\beta_2\dots}
+\cdots + w  \O^{\alpha_1\alpha_2\dots}{}_{\beta_1\beta_2\dots} \p_\mu \zeta^\mu \ . \notag
\end{align}
The replacement $\p_\mu \to \del_\mu$ leaves this expression unchanged. 
For a tensor density $\tO$ of weight one, we can write $\tO = \sqrt{g} \O$ where $\O$ has weight zero, and the Lie derivative satisfies (suppressing indices on $\O$, which may have spin)
\begin{align}
{\cal L}_\zeta \tilde{\O} = \sqrt{g}{\cal L}_\zeta \O + \sqrt{g} \O \del_\mu \zeta^\mu\, .
\end{align}
For the stress tensor, the explicit formulae are
\begin{align}\label{LieT}
{\cal L}_\zeta T_{\alpha\beta} &= \zeta^\mu \p_\mu T_{\alpha\beta} +  \p_\alpha \zeta^\mu T_{\mu\beta} + \p_\beta \zeta^\mu T_{\alpha \mu} \\
{\cal L}_\zeta \tT_{\alpha\beta} &= \sqrt{g} {\cal L}_\zeta T_{\alpha\beta}  + \sqrt{g} T_{\alpha\beta} \del_\mu \zeta^\mu \, .
\end{align}
We now turn to the Ward identities. Recall that the metric transforms under diffeomorphisms as ${\cal L}_\zeta g^{\mu\nu} = -2 \del^{(\mu} \zeta^{\nu)}$. Diffeomorphism invariance of the effective action $\log Z[g]$ requires
\begin{align}\label{zerodZ}
0 &=\delta_{\zeta} \log Z =  \int d^d x \left({\cal L}_\zeta g^{\mu\nu}\right) \frac{\delta}{\delta g^{\mu\nu}} \log Z
%
= - \int d^d x \sqrt{g} \zeta^\nu \del^\mu \langle T_{\mu\nu}\rangle\, .
\end{align}
This implies the conservation of stress energy in a general background, $\del^\mu \langle T_{\mu\nu}\rangle = 0$. For a correlation function, coordinate invariance under $x \to x'$ requires
\begin{align}
\langle \O'(x_1)\O'(x_2)\cdots\rangle_g = \langle \O(x_1)\O(x_2)\cdots\rangle_{g'}\, .
\end{align}
where $g$ is the background metric. The infinitessimal form of this equation is
\begin{align}\label{genDiffL}
 \langle  {\cal L}_\zeta \O(x_1) \O(x_2)\cdots\rangle
  + &\langle \O(x_1) {\cal L}_\zeta \O(x_2)\cdots\rangle + \cdots
  = \delta_\zeta \langle \O(x_1)\O(x_2)\cdots\rangle\\
  &= \int d^d x{\cal L}_\zeta g^{\mu\nu} \frac{\delta}{\delta g^{\mu\nu}} \langle \O(x_1)\O(x_2)\cdots\rangle\notag\, ,
\end{align}
where $\delta_\zeta$ acts only on the background metric, and $\O$ is a tensor density of any rank. We can take $\O = \tT_{\alpha\beta}$, but not $T_{\alpha\beta}$ for the reasons above stemming from the Osborn-Petkou conventions. Each operator insertion transforms independently, so without loss of generality we can focus on the one-point function. Then \eqref{genDiffL} becomes
\begin{align}\label{tdward}
\int d^d x \sqrt{g} \zeta^\nu(x) \del^\mu \langle T_{\mu\nu}(x) \tT_{\alpha\beta}(x_1)\rangle
&=- {\cal L}_\zeta \langle \tT_{\alpha\beta}(x_1)\rangle\, .
\end{align}
Dividing by $\sqrt{g(x_1)}$ in \eqref{tdward} to write this in terms of the Osborn-Petkou stress tensor and using \eqref{LieT} gives
\begin{align}\label{tward}
\int d^d x \sqrt{g} \zeta^\nu \del^\mu \langle T_{\mu\nu}(x) T_{\alpha\beta}(x_1) \rangle
&= -{\cal L}_\zeta \langle T_{\alpha\beta}(x_1)\rangle - \langle T_{\alpha\beta}(x_1)\rangle \del_\mu \zeta^\mu(x_1) \ .
\end{align}
This can also be derived by varying the conservation equation $\del^\mu\langle T_{\mu\nu}\rangle = 0$; see section \ref{ss:varycons} below. The extra term on the right indicates that the Osborn-Petkou stress tensor  transforms as a weight-one tensor density inside correlation functions.  Stripping off the integral gives the conservation law
\begin{align}
\del^\mu \langle T_{\mu\nu}(x) T_{\alpha\beta}(x_1)\rangle
&=
\langle T_{\alpha\beta}(x)\rangle \del_\nu \delta^{(d)}(x-x_1)
+
\del_\alpha \left( \langle T_{\nu\beta}(x)\rangle \delta^{(d)}(x-x_1) \right)\\
&\qquad 
+\del_\beta \left( \langle T_{\alpha\nu}(x)\rangle \delta^{(d)}(x-x_1) \right)\notag\, .
\end{align}
In $n$-point functions of $T_{\mu\nu}$, we get a sum of such terms for each insertion. These expressions agree with \cite{Osborn:1993cr}.

Now consider correlation functions involving $\Theta$ insertions. Recall from \eqref{ThetaDefCor} that correlators of $\Theta$ are defined with the trace inside the variation, but the $1/\sqrt{g}$ outside. It follows that $\Theta$ behaves like a weight one scalar density, i.e., 
\begin{align}
\int d^d x \sqrt{g} \zeta^\nu \del^\mu\langle T_{\mu\nu}(x) \Theta(x_1)\rangle
&= -{\cal L}_\zeta \langle \Theta(x_1)\rangle - \langle \Theta(x_1)\rangle \del_\mu \zeta^\mu(x_1) \ .
\end{align}
Using ${\cal L}_\zeta \Theta = \zeta^\alpha \del_\alpha \Theta$ and stripping off the integral this implies
\begin{align}\label{apward1}
\del^\mu \langle T_{\mu\nu}(x)\Theta(x_1)\rangle &= 
\langle \Theta(x)\rangle \del_\nu \delta^{(d)}(x-x_1) \ .
\end{align}
For the 3-point function,
\begin{align}\label{apward2}
\del^\mu \langle T_{\mu\nu}(x)\Theta(x_1)\Theta(x_2)\rangle &= 
\langle \Theta(x)\Theta(x_2)\rangle \del_\nu \delta^{(d)}(x-x_1) 
+\langle \Theta(x)\Theta(x_1)\rangle \del_\nu \delta^{(d)}(x-x_2) 
\ .
\end{align}
These are superficially different from the standard Ward identity for a scalar of dimension $d$, which would have $-\del_\nu \langle \O\rangle \delta^{(d)}$ on the right-hand side (see e.g. \cite{DiFrancesco:1997nk}). In those references, the operators are implicitly defined with the $1/\sqrt{g}$'s inside the variations.

\subsection{Symmetries}

Given a closed codimension-1 surface $\Sigma = \p V$, define the surface deformation acting on a local operator $\O(x_1)$ by
\begin{align}\label{surfaceDef}
S[\zeta]\circ \O(x_1) &= \oint d\Sigma^\mu \zeta^\nu T_{\mu\nu} \O(x_1) \notag \\
&= -\int_V \sqrt{g} \del^\mu \zeta^\nu T_{\mu\nu} \O(x_1) -
\int_V \sqrt{g} \zeta^\nu \del^\mu T_{\mu\nu} \O(x_1)\, .
\end{align}
If $\O$ is a true tensor, e.g. $\O = \tT_{\alpha\beta}$ or $\O = \tTheta$, then assuming $\Sigma$ encloses $x_1$, the last term is $-{\cal L}_\zeta \langle \O(x_1)\rangle$. For $\O = T_{\alpha\beta}$ or $\Theta$ there is an extra term from \eqref{tward}. 

 If $\zeta$ generates a symmetry, i.e. $\del^\mu \zeta^\nu T_{\mu\nu} = 0$ up to contact terms, then $S[\zeta]$ is a topological operator. By slicing $\Sigma$ in half in the standard way we can reinterpret its action as the commutator
\begin{align}
S[\zeta]\circ \O(x_1) = [Q_\zeta , \O(x_1)] \ .
\end{align}
For the stress tensor, accounting for the extra term in \eqref{tward}, this implies
\begin{align}\label{gencom}
[Q_\zeta, T_{\alpha\beta}(x_1)] &= 
 {\cal L}_\zeta T_{\alpha\beta}(x_1) + T_{\alpha\beta}(x_1) \del_\mu \zeta^\mu(x_1)
- \int_{x\sim x_1} \sqrt{g(x)} \del^\mu \zeta^\nu(x) T_{\mu\nu}(x) T_{\alpha\beta}(x_1)\, .
\end{align}
We now set $g_{\mu\nu} = \delta_{\mu\nu}$. The isometries of the flat metric yield the Poincare Ward identities, for which the last term in \eqref{gencom} vanishes using $\p^{(\mu} \zeta^{\nu)} = 0$.

\subsection{Dilatations in CFT}

If the theory is conformal, then there is a dilatation symmetry $\zeta^\alpha = x^\alpha$ with conserved charge $D$. The action of the dilatation on the stress tensor in $d$ dimensions is
\begin{align}\label{canonicalT}
[D, T_{\alpha\beta}] = x^\mu \p_{\mu} T_{\alpha\beta} + d\, T_{\alpha\beta} \ .
\end{align}
For the conformal Ward identities, unlike Poincare, the last term in \eqref{gencom} is important. Dilatations have  $\p^{(\alpha} \zeta^{\beta)} = g^{\alpha\beta}$, with $g^{\alpha\beta}$ the flat metric. Denote the contact term in $T_{\mu\nu}(x)\O(x_1)$ by brackets, $\{ T_{\mu\nu} \O \}$. Consistency of \eqref{gencom} with \eqref{canonicalT} requires the last term in \eqref{gencom} to contribute
\begin{align}
\int_{x \sim x_1} d^d x \sqrt{g(x)} g^{\mu\nu}(x) \{ T_{\mu\nu}(x) T_{\alpha\beta}(x_1) \} = 2 T_{\alpha\beta}(x_1) \ .
\end{align}
Therefore
\begin{align}\label{cftTraceContact}
 g^{\mu\nu}(x)\{ T_{\mu\nu}(x) T_{\alpha\beta}(x_1) \} = 2 T_{\alpha\beta}(x_1) \delta^{(d)}(x-x_1) + \mbox{total $x$-derivatives} \ .
\end{align}
A change in conventions  as to whether the variation is done with respect to $g^{\alpha\beta}$ or $g_{\alpha\beta}$, and where the $1/\sqrt{g}$'s are placed, would not alter \eqref{canonicalT}, but it would change the contact term \eqref{cftTraceContact}. The effect of such a change in conventions is to move contributions between the two terms in \eqref{surfaceDef}.

This discussion and in particular \eqref{cftTraceContact} apply in all spacetime dimensions, with and without a conformal anomaly. (We have assumed there is no diff anomaly.) In odd dimensions where there is no conformal anomaly, \eqref{cftTraceContact} simply comes from expanding out the metric variations in the equation $\frac{\delta}{\delta g^{\alpha\beta}(x_1)} \langle T^{\mu}_\mu(x) \cdots \rangle = 0$. In even dimensions, \eqref{cftTraceContact} indicates that the anomaly cannot alter the $\{ T^\mu_\mu T_{\alpha\beta} \} \to 2T_{\alpha\beta}$ contact term. The only effect of the anomaly here is to add a $c$-number total derivative in 
\eqref{cftTraceContact} that is responsible for the nonzero two-point function $\langle T^{\mu}_\mu(x_1)T_{\alpha\beta}(x_2)\rangle$, proportional to $c$ in both two and four dimensions.

Next let us repeat this analysis for a local scalar primary operator, $\O$, in a CFT. Define correlation functions by adding a source  $\int \sqrt{g} J \O$ and varying as
\begin{align}
\langle \O\rangle &= \frac{1}{\sqrt{g}} \frac{\delta }{\delta J} \log Z \\
\langle T_{\alpha\beta}(x_1) \O(x_2)\rangle
&= \frac{-2}{\sqrt{g(x_1)} \sqrt{g(x_2)}} \frac{\delta}{\delta g^{\alpha\beta}(x_1)} \frac{\delta}{\delta J(x_2)} \log Z\, ,
\end{align}
etc. The action of the dilatation is
$
[D, \O] = x^\mu \p_\mu \O + \Delta \O
$, and the conservation equation analogous to \eqref{gencom} is
\begin{align}
[D, \O(x_1)] = \p_{x_1^\mu} (x_1^\mu \O(x_1)) - \int_{x\sim x_1} \sqrt{g} g^{\mu\nu}T_{\mu\nu} \O(x_1) \ .
\end{align}
Therefore the contact term is
\begin{align}
g^{\mu\nu}(x)\{ T_{\mu\nu}(x) \O(x_1) \} &= (d-\Delta) \O(x_1) \delta^{(d)}(x-x_1)  + \mbox{total $x$-derivatives} \ .
\end{align}
This agrees with \cite{Osborn:1993cr}.

Finally, consider the operator $\Theta$, defined in \eqref{ThetaDefCor}. This operator behaves under diffeomorphisms and dilatations exactly like a scalar primary of dimension $d$. Therefore the contact term vanishes upon integration:
\begin{align}\label{thetacontactapp}
g^{\mu\nu}(x)\{T_{\mu\nu}(x) \Theta(x_1) \} &= \mbox{total $x$-derivatives} \ .
\end{align}
This can also be derived by taking the trace of \eqref{cftTraceContact} and using 
$
g^{\mu\nu}(x_1) \frac{\delta}{\delta g_1^{\mu\nu}(x)} g^{\alpha\beta}(x_2) \frac{\delta}{\delta g^{\alpha\beta}(x_2)} = 
g^{\mu\nu}(x_1)g^{\alpha\beta}(x_2)  \frac{\delta}{\delta g_1^{\mu\nu}(x)} \frac{\delta}{\delta g^{\alpha\beta}(x_2)}
 + \sqrt{g} \delta^{(d)}(x_1-x_2) g^{\mu\nu} \frac{\delta} {\delta g_{\mu\nu}}$.



\subsection{Varying the conservation equation}\label{ss:varycons}
The Ward identities \eqref{apward1}-\eqref{apward2} can also be derived by varying the conservation equation, $\del^\mu\langle T_{\mu\nu}\rangle =0 $, with respect to the background metric. This is a useful check of the signs.

To ease the notation we adopt the convention that all quantities are evaluated at the point $x$, unless there is a sub/superscript, so $T_{\mu\nu} = T_{\mu\nu}(x)$, $T_{\mu\nu}^{(1)} = T_{\mu\nu}(x_1)$, $\Theta_2 = \Theta(x_2)$, etc. Delta functions are written $\delta_i = \delta^{(d)}(x-x_i)$ and $\delta_{ij} = \delta^{(d)}(x_i - x_j)$. All derivatives are with respect to $x$. Everything in this subsection is in Euclidean signature. Define the variations
\begin{align}
\Delta_{\mu\nu} &= -\frac{2}{\sqrt{g}} \frac{\delta}{\delta g^{\mu\nu}} \ , \qquad & \Delta & = g^{\mu\nu}\Delta_{\mu\nu}\, ,\\
\Delta^{(i)}_{\mu\nu} &=- \frac{2}{\sqrt{g(x_i)}} \frac{\delta}{\delta g^{\mu\nu}(x_i)} , \qquad &
\Delta^{(i)} & = g^{\mu\nu}(x_i) \Delta^{(i)}_{\mu\nu} \ .  \notag
\end{align}
The correlators are
\begin{align}
\langle T_{\mu\nu}\rangle &= \Delta_{\mu\nu} \log Z \\
\langle T_{\mu\nu}\Theta_1\rangle &= \frac{1}{\sqrt{g_1}} \Delta_{\mu\nu} \sqrt{g_1}\Delta^{(1)}\log Z \notag\\
\langle T_{\mu\nu}\Theta_1\Theta_2\rangle &= \frac{1}{\sqrt{g_1}\sqrt{g_2}} \Delta_{\mu\nu} \sqrt{g_1}\Delta^{(1)} \sqrt{g_2}\Delta^{(2)} \log Z \, .\notag
\end{align}
A useful identity for the Weyl variation of a covariant derivative is
\begin{align}\label{handy1}
\Delta^{(1)} \del^\mu Y_{\mu\nu} &= \del^{\mu}  \Delta^{(1)} Y_{\mu\nu}-2\delta_1\del^\mu Y_{\mu\nu}  + (d-2)Y_{\mu\nu}\del^\mu \delta_1 - Y^\alpha_\alpha \del_\nu \delta_1 \, ,
\end{align}
where $\phi$ is a scalar and $Y_{\mu\nu}$ is a symmetric tensor. The Weyl variation of the 1-point function is 
\begin{align}
\Delta^{(1)}\langle T_{\mu\nu}\rangle
&= \frac{1}{\sqrt{g_1}} \sqrt{g_1}\Delta^{(1)} \Delta_{\mu\nu}\log Z\\
&=\langle T_{\mu\nu}\Theta_1\rangle + \frac{1}{\sqrt{g_1}} [ \sqrt{g_1}\Delta^{(1)}, \Delta_{\mu\nu}]\log Z \notag\, .
\end{align}
A short calculation gives the commutator
$
[\sqrt{g_1}\Delta^{(1)}, \Delta_{\mu\nu}]  = -(d-2) \delta_1 \sqrt{g} \Delta_{\mu\nu}
$.
Therefore
\begin{align}\label{weylT}
\Delta^{(1)}\langle T_{\mu\nu}\rangle 
&=\langle T_{\mu\nu}\Theta_1\rangle  - (d-2)\delta_1 \langle T_{\mu\nu}\rangle\, .
\end{align}
Now to get the first order Ward identity we vary the conservation equation:
\begin{align}
0 &= \Delta^{(1)} \del^\mu \langle T_{\mu\nu}\rangle  \\
&= \del^\mu \Delta^{(1)} \langle T_{\mu\nu}\rangle
-2\delta_1 \del^\mu \braket{T_{\mu\nu}} + (d-2) \langle T_{\mu\nu}\rangle \del^\mu \delta_1 - \langle T^{\alpha}_\alpha \rangle \del_\nu \delta_1 \notag\\
 &=  \del^\mu \langle T_{\mu\nu}\Theta_1\rangle - \langle \Theta \rangle \del_\nu \delta_1 \notag\, .
\end{align}
We used \eqref{handy1} then \eqref{weylT}. Thus we have the first Ward identity,
\begin{align}\label{ward1app}
\del^\mu\langle T_{\mu\nu}\Theta_1\rangle 
= \langle \Theta\rangle \del_\nu \delta_1\, ,
\end{align}
in agreement with \eqref{apward1}. Varying this equation again under $\Delta^{(2)}$, the Weyl variation at the point $x_2$, gives \eqref{apward2}.

\section{Details of CFT calculations\label{ap:FourDim}}

In this appendix, we calculate the Euclidean two- and three-point functions $\langle  T_{\mu\nu} \Theta\rangle$ and $\langle T_{\mu\nu}\Theta\Theta\rangle$ in 4d CFT by varying the trace anomaly. The results are used in section \ref{ss:cftLocalContact} in the main text.

\subsection{Trace anomaly}

In a four-dimensional CFT on a curved background, the trace is
\be 
\braket{\Theta(x)} = -a E_4+ c W_{\mu\nu\rho\sigma}^2 + b_1 \square R + b_2 \Lambda^2 R + b_3 \Lambda^4\, ,\label{eq:Trace4dAP}
\ee
where $E_4$ is the Euler density and $W_{\mu\nu\rho\sigma}$ is the Weyl tensor. As explained under \eqref{theta4d}, it is possible (and customary) to set $b_1=b_2=b_3=0$ by adding local counterterms. In even dimensions, the Euler density is defined as 
\be 
E_{2n} = \frac{1}{2^n}R_{\mu_1\nu_1\rho_1\sigma_1}\cdots R_{\mu_n\nu_n\rho_n\sigma_n}\epsilon^{\mu_1\nu_1\mu_2j_2\dots \mu_n\nu_n}\epsilon^{\rho_1\sigma_1\rho_2\sigma_2\dots \rho_n\sigma_n}\, .
\ee
The Weyl tensor is given by 
\be 
W_{\mu\nu\rho\sigma} = R_{\mu\nu\rho\sigma} + \frac{1}{d-2}(g_{\mu\sigma}R_{\nu\rho}-g_{\mu\rho}R_{\nu\sigma} + g_{\nu\rho}R_{\mu\sigma}-g_{\nu\sigma}R_{\mu\rho})+\frac{1}{(d-2)(d-1)}(g_{\mu\rho}g_{\nu\sigma}-g_{\mu\sigma}g_{\nu\rho})R\, .
\ee
In four dimensions, and using all the symmetries of the Riemann tensor, we have
\begin{align}
E_4 &= R_{\mu\nu\rho\sigma}^2-4R_{\mu\nu}^2+R^2\, ,\\
W_{\mu\nu\rho\sigma}^2 &= R_{\mu\nu\rho\sigma}^2 -2R_{\mu\nu}^2 +\frac{1}{3}R^2\, ,
\end{align}
with $R_{\mu\nu\rho\sigma}$ the Riemann tensor, $R_{\mu\nu}$ the Ricci tensor and $R$ the Ricci scalar. In terms of these tensors, the trace anomaly \eqref{eq:Trace4dAP} is
\be 
\braket{\Theta} =(c-a)R_{\mu\nu\rho\sigma}^2+ 2(2a-c)R_{\mu\nu}^2  + \left(\frac{c}{3}-a\right)R^2 + b_1 \nabla^\mu\nabla_\mu R  + b_2 \Lambda^2 R + b_3\Lambda^4\, .\label{eq:TraceAnom4dp}
\ee

\subsection{$\langle\Theta\Theta\rangle$}

Applying the definition \eqref{ThetaDefCor}, the Euclidean 2-point function is given by the variation
\begin{align}
\langle\Theta(x_1)\Theta(x_2)\rangle &= 
-\frac{2}{\sqrt{g(x_1)} \sqrt{g(x_2)}}  g^{\alpha\beta}(x_2) \frac{\delta }{\delta g^{\alpha\beta}(x_2)}\left( \sqrt{g(x_1)} \langle\Theta(x_1)\rangle \right)  \ .
\end{align}
Varying \eqref{eq:Trace4dAP} in a general curved background gives
%
\begin{align}\label{ThetaThetaCurved}
&\braket{\Theta(x_1)\Theta(x_2)}  = \\
&\quad  \Big[ -8aG_{\mu\nu}\nabla^\mu \nabla^\nu -2b_1\left(R\nabla^2  +\nabla^\mu R\nabla_\mu + 3 \nabla^4\right)+2b_2\Lambda^2 (R - 3\nabla^2)+4b_3\Lambda^4\Big]\delta^{(4)}(x_{12})\, ,\nonumber 
\end{align}
where $x_{ij} \equiv x_i-x_j$, 
and 
 $G_{\mu\nu} = R_{\mu\nu}-\frac{1}{2}g_{\mu\nu}R$ is the Einstein tensor. In flat Euclidean space,
\begin{align}
&\braket{\Theta(x_1)\Theta(x_2)} = \Big[-6b_2\Lambda^2 \partial^2   -6b_1 \partial^4+4b_3\Lambda^4\Big]\delta^{(4)}(x_{12})\, .\nonumber 
\end{align}
The Fourier transform to Euclidean momentum space gives
\begin{align}
\lll \Theta(\K_1)\Theta(-\K_1)\rr=& -6b_1 \K_1^4+ 6 b_2 \Lambda^2\K_1^2+4b_3\Lambda^4 \, .\label{eq:ThetaTheta4dAP}
\end{align}

\subsection{$\langle T_{\mu\nu}\Theta\rangle$}

Using the definition in \eqref{mixedDefs}, 
\begin{align}
\braket{\Theta(x_1)T_{\mu\nu}(x_2)}  =&  -\frac{2}{\sqrt{g(x_1)}\sqrt{g(x_2)}}\frac{\delta}{\delta g^{\mu\nu}(x_2)}\left[\sqrt{g(x_1)}\braket{\Theta(x_1)}\right]\, .\label{eq:Euclidean4dTT}
\end{align}
Performing the variation, we obtain (in flat Euclidean space)
\begin{align}
\braket{\Theta(x_1)T_{\mu\nu}(x_2)} =& \left[2b_1 (\partial_\mu \partial_\nu \partial^2 -g_{\mu\nu}\partial^4)+ 2b_2\Lambda^2( \partial_\mu \partial_\nu - g_{\mu\nu}\partial^2)+b_3 \Lambda^4 g_{\mu\nu} \right]\delta^{(4)}(x_{12})\, , \label{eq:Euclidean4dTT2}
\end{align}
with $\partial^2 = \partial_\mu \partial^\mu$. 
In Euclidean momentum space,
\begin{align}
\lll\Theta(\K_1)T_{\mu\nu}(-\K_1)\rr &= 2b_1\K_1^2 (\K_{1\mu}\K_{1\nu} -g_{\mu\nu}\K_1^2)-2b_2\Lambda^2( \K_{1 \mu}\K_{1\nu}- g_{\mu\nu}\K_1^2)+b_3 \Lambda^4 g_{\mu\nu} \, , \label{eq:Euclidean4dTT2k}
\end{align}
whose trace agrees with \eqref{eq:ThetaTheta4dAP}. The Ward identity \eqref{apward1} is also satisfied,
\begin{align}
\K^\mu \lll \Theta(-\K) T_{\mu\nu}(\K) \rr
&= \K_{\nu} \langle \Theta \rangle_{g_{\mu\nu} = \delta_{\mu\nu}} \ , 
\end{align}
since the trace in flat space is $\langle \Theta \rangle_{g_{\mu\nu} = \delta_{\mu\nu}} = b_3 \Lambda^4$.

\subsection{$\langle T_{\mu\nu}\Theta\Theta\rangle$}
The 3-point function is cumbersome but straightforward. With the definition in \eqref{mixedDefs} it is given by the variation
\be 
\braket{\Theta(x_1)\Theta(x_2)T_{\mu\nu}(x_3)}  = -\frac{2}{\sqrt{g(x_1)}\sqrt{g(x_2)}\sqrt{g(x_3)}}\frac{\delta}{\delta g^{\mu\nu}(x_3)}\left[\sqrt{g(x_1)}\sqrt{g(x_2)}\braket{\Theta(x_1)\Theta(x_2)}\right]\, .\label{eq:ToVary1AP}
\ee
The $\langle\Theta\Theta\rangle$ 2-point function in a general background is \eqref{ThetaThetaCurved}. Note that the covariant delta function has a nontrivial metric variation,
\be 
\delta \left[\delta^{(d)}(x_{12}) \right]= \frac{1}{2} \delta^{(d)}(x_{12}) g_{\alpha\beta}\delta g^{\alpha\beta} \ .
\ee
%
%
Varying \eqref{ThetaThetaCurved} once more we obtain the Euclidean correlator (in flat space),
%
\begin{align}
&\braket{T_{\mu\nu}(x_3)\Theta(x_1)\Theta(x_2)} = 4 b_3 \Lambda^4 g_{\mu\nu}\delta_{12}\delta_{13}\label{eq:3ptposmess}\\
&\quad -b_2 \Lambda^2g_{\mu\nu}\Big[6 \partial_\alpha \delta_{12}\partial^\alpha \delta_{13}+4 \delta_{12}\partial^2 \delta_{13}+6 \partial^2 \delta_{12}\delta_{13}\Big]\nonumber\\
&\quad +b_2\Lambda^2 \Big[12\partial_{(\mu}\delta_{12}\partial_{\nu)}\delta_{13} + 12 \partial_\nu \partial_\mu \delta_{12}\delta_{13} +4 \delta_{12}\partial_\nu \partial_\mu \delta_{13}\Big]\nonumber\\
&\quad + 2b_1 g_{\mu\nu}\Big[2 \partial^2 \delta_{12}\partial^2 \delta_{13}-6 \partial_\alpha \partial^2 \delta_{12}\partial^\alpha \delta_{13}-\partial_\alpha \delta_{12}\partial^\alpha \partial^2 \delta_{13}-3\partial^4\delta_{12}\delta_{13}-6 \partial_\alpha \partial_\beta \delta_{12}\partial^\alpha \partial^\beta \delta_{13}\Big]\nonumber\\
&\quad + b_1\Big[12\partial_{(\mu} \delta_{12}\partial_{\nu)} \partial^2 \delta_{13} + 24 \partial_{(\mu} \partial^2 \delta_{12}\partial_{\nu)} \delta_{13} + 24 \partial_{(\mu} \partial^\alpha \delta_{12}\partial_{\nu)} \partial_\alpha \delta_{13}+ 12 \partial_\nu\partial_\mu \delta_{12}\partial^2\delta_{13}\Big]\nonumber\\
&\quad + b_1\Big[24 \partial_{\nu}\partial_\mu \partial_\alpha \delta_{12}\partial^\alpha \delta_{13}-4\partial^2 \delta_{12}\partial_{\nu}\partial_{\mu}\delta_{13}  - 4 \partial^\alpha \delta_{12}\partial_\nu \partial_\mu \partial_\alpha \delta_{13} + 24 \partial_\nu\partial_\mu \partial^2 \delta_{12}\delta_{13}\Big]\nonumber\\
&\quad + ag_{\mu\nu}\Big[-8 \partial^2 \delta_{12}\partial^2 \delta_{13} + 8 \partial_\alpha \partial_\beta \delta_{12}\partial^\alpha\partial^\beta \delta_{13}\Big]\nonumber\\
&\quad +a\Big[ 8 \partial_\nu \partial_\mu \delta_{12}\partial^2\delta_{13} + 8\partial^2 \delta_{12}\partial_\nu \partial_\mu \delta_{23}-16 \partial^\alpha\partial_{(\mu} \delta_{12}\partial_{\nu)} \partial_\alpha \delta_{13}\Big]\, ,\nonumber
\end{align}
where we defined $\delta_{ij}\equiv \delta^{(4)}(x_i-x_j)$, and all derivatives are with respect to $x_1$. 
It is trivial to transform to momentum space by first converting $x_1$-derivatives to derivatives with respect to $x_2$ and $x_3$, so for example $\p_\mu \p_\nu \delta_{12} \p^2 \delta_{13} \to \K_{2\mu} \K_{2\nu} \K_3^2$, etc. The result is the Euclidean momentum space correlator
\begin{align}
&\lll T_{\mu\nu}(\K_3)\Theta(\K_1)\Theta(\K_2)\rr =  4b_3\Lambda^4g_{\mu\nu}\nonumber\\
&\quad + 4b_2\Lambda^2\Big[g_{\mu\nu}(\K_1^2 +\K_2^2 + \frac{1}{2}\K_1\cdot \K_2)-\K_{1\mu}\K_{1\nu}-\K_{2\mu}\K_{2\nu}+ \K_{1(\mu}\K_{2\nu)}\Big]\nonumber\\
&\quad + 2b_1\Big[g_{\mu\nu}\Big(-4(\K_1\cdot \K_2)^2 + 3 \K_1^2 \K_2^2 + \K_1\cdot \K_2 (\K_1^2 + \K_2^2) \Big)\Big]\nonumber\\
&\quad + 2b_1 \Big[2\K_1\cdot \K_2 (\K_1+\K_2)_\mu(\K_1+\K_2)_\nu-6 \K_{1(\mu}\K_{2\nu)}(\K_1^2 + \K_2^2)\Big]\nonumber\\
&\quad+ 8a\Big[g_{\mu\nu}\left((\K_1\cdot \K_2)^2 -\K_1^2\K_2^2\right) + \K_{1\mu}\K_{1\nu} \K_2^2 + \K_{2\mu}\K_{2\nu}\K_1^2 - 2(\K_1\cdot \K_2)\K_{1(\mu}\K_{2\nu)}\Big]\, .\label{eq:Tuvthetatheta4dap}
\end{align}
This satisfies the Ward identity \eqref{apward2}, which in momentum space requires
\begin{align}
\K_3^\mu \lll T_{\mu\nu}(\K_3)\Theta(\K_1)\Theta(\K_2)\rr 
&= -\K_{1\nu} \lll \Theta(\K_2)\Theta(-\K_2) \rr
 - \K_{2\nu} \lll \Theta(\K_1)\Theta(-\K_1)\rr
\end{align}
for $\K_1 + \K_2 + \K_3 = 0$, with $\langle \Theta\Theta\rangle$ from \eqref{eq:ThetaTheta4dAP}. 

The null-null component of the 3-point function, in Euclidean signature, is
\begin{align}
\lll T_{uu}(\K_3)\Theta(\K_1)\Theta(\K_2)\rr &=  -4b_2\Lambda^2\Big[\K_{1u}^2+\K_{2u}^2- \K_{1u}\K_{2u}\Big]\nonumber\\
&\quad + 4b_1 \Big[\K_1\cdot \K_2 (\K_{1u}+\K_{2u})(\K_{1u}+\K_{2u})-3 \K_{1u}\K_{2u}(\K_1^2 + \K_2^2)\Big]\nonumber\\
&\quad + 8a\Big[  \K_{1u}^2 \K_2^2 + \K_{2u}^2\K_1^2 - 2(\K_1\cdot \K_2)\K_{1u}\K_{2u}\Big]\, .
\end{align}

\section{Massive scalar computations}\label{app:loops}

In this appendix, we calculate a few 1-loop diagrams for the massive scalar theory in dimensional regularization. The results here are not used in the main text, except as a second way to calculate the diagram in figure \ref{fig:TuuThetaTheta}. The correlators written in this appendix are regulated by not renormalized. We work in Euclidean spacetime, where the propagator is
\be 
G_{ij}\equiv G(x_i-x_j) = \int \frac{d^dp}{(2\pi)^d}\frac{e^{ip\cdot (x_{i}-x_j)}}{p^2 + m^2}\, .\label{eq:D3}
\ee
We will frequently use $x_{ij}\equiv x_i-x_j$ as well as the following standard loop integral results
\begin{align}
\int \frac{d^dp}{(2\pi)^d}\frac{1}{(p^2 + \Delta)^n}& = \frac{1}{(4\pi)^{d/2}}\frac{\Gamma(n-\frac{d}{2})}{\Gamma(n)}\frac{1}{\Delta^{n-\frac{d}{2}}}\label{eq:D4}\, ,\\
\int \frac{d^dp}{(2\pi)^d}\frac{p_\mu p_\nu}{(p^2 + \Delta)^n}& = \frac{1}{(4\pi)^{d/2}}\frac{g_{\mu\nu}}{2}\frac{\Gamma(n-1-\frac{d}{2})}{\Gamma(n)}\frac{1}{\Delta^{n-1-\frac{d}{2}}}\, ,\\
\int \frac{d^dp}{(2\pi)^d}\frac{p_\mu p_\nu p_\rho p_\sigma }{(p^2 + \Delta)^n}& = \frac{1}{(4\pi)^{d/2}}\frac{1}{4}\left(g_{\mu\nu}g_{\rho\sigma}+g_{\mu\rho}g_{\nu\sigma}+g_{\mu\sigma}g_{\nu\rho}\right)\frac{\Gamma(n-2-\frac{d}{2})}{\Gamma(n)}\frac{1}{\Delta^{n-2-\frac{d}{2}}}\label{eq:D6}\, .
\end{align}
The stress tensor in four dimensions can be found in \eqref{eq:TarbitraryD} (and the trace in \eqref{eq:TracearbitraryD}). Moreover, as explained in the main text, we can set $\Theta$ to $-m^2\phi^2$ up to contact terms.

%
%

\subsection{$\langle \Theta\Theta\rangle$}
The trace-trace two-point function computation is a simple Wick contraction exercise. The loop integral that we ultimately need to evaluate is
\begin{align}
\lll\Theta(\K_1)\Theta(-K_1)\rr
& = \int\frac{d^4p}{(2\pi)^4}\frac{2m^4}{(p^2+m^2)((p+\K_1)^2+m^2)}\, .
\end{align}
We perform this integral using Feynman parametrization 
\be 
\Delta = m^2 +\K_1^2 x(1-x), \qquad \qquad p\rightarrow p-x\K_1\, ,\label{eq:D12}
\ee
with $x$ the Feynman parameter and the loop results \eqref{eq:D4}-\eqref{eq:D6}. The $O(\K^4)$ term is
\begin{align}
\left.\lll\Theta(\K_1)\Theta(-\K_1)\rr\right|_{O(\K^4)}
=\frac{\K_1^4}{480 \pi
   ^2} \label{eq:TraceTrace4d}\, .
\end{align}


\subsection{$\langle T_{\mu\nu}\Theta\rangle$}
We now compute the two-point function $\braket{T_{\mu\nu}(x_1)\Theta(x_2)}$. After doing the Wick contractions, this two-point function in position space is given by 
\begin{align}
\braket{T_{\mu\nu}(x_1)\Theta(x_2)} &= -\frac{4m^2}{3} \partial_{1\mu}G_{12}\partial_{1\nu}G_{12}+\frac{2m^2}{3}\left(G_{12}\partial_{1\mu}\partial_{1\nu}G_{12}  -g_{\mu\nu}G_{12}\square_1G_{12}\right)\nonumber\\
&\phantom{=}+g_{\mu\nu}m^4G_{12}^2 +\frac{1}{3} m^2 g_{\mu\nu}\partial_{1\alpha}G_{12}\partial_1^\alpha G_{12}\, ,
\end{align}
which is 
\begin{align}
\braket{T_{\mu\nu}(x_1)\Theta(x_2)} &= m^2\int\frac{d^4p_1}{(2\pi)^4}\frac{d^4p_2}{(2\pi)^4}\frac{e^{i(p_1+p_2)\cdot x_{12}}(\frac{4}{3}p_{1\mu}p_{2\nu}-\frac{2}{3}p_{2\mu}p_{2\nu}+g_{\mu\nu}(\frac{2}{3}p_2^2-\frac{1}{3}p_1\cdot p_2+ m^2) )}{(p_1^2+m^2)(p_2^2+m^2)}\, .
\end{align}
In momentum space, we obtain
\begin{align}
\lll T_{\mu\nu}(\K_1)\Theta(-\K_1)\rr &= m^2\int\frac{d^4p}{(2\pi)^4}\frac{-\frac{2}{3}(3p_{\mu}+ 2\K_{1\mu})p_{\nu}+g_{\mu\nu}(\frac{1}{3}p\cdot \K_1 + p^2 +m^2)}{(p^2+m^2)((p+\K_1)^2+m^2)}\, .\nonumber
\end{align}
Performing the integral with the same Feynman parameter, we finally obtain
\begin{align} 
\left. \lll T_{\mu\nu}(\K_1)\Theta(-\K_1)\rr\right|_{O(\K^4)} &=-\frac{ \left(\K_{1\mu}\K_{1\nu}-g_{\mu\nu}\K_1^2\right)}{1440\pi^2}\K_1^2\, .
\end{align}
We can take the trace, and obtain 
\begin{align} 
\left.g^{\mu\nu}\lll T_{\mu\nu}(x_1)\Theta(x_2)\rr\right|_{O(\K^4)} &= \frac{\K_1^4}{480\pi^2}\, ,
\end{align}
which matches with \eqref{eq:TraceTrace4d}.

\subsubsection{$\langle T_{\mu\nu}\Theta\Theta\rangle$}
The three-point function is given by 
\begin{align}
\braket{T_{\mu\nu}(x_3)\Theta(x_1)\Theta(x_2)} &=\frac{4m^4}{3}G_{12}\Big[2\left(\partial_{2\mu}\partial_{1\nu}+ \partial_{1\mu}\partial_{2\nu}\right)-\Big(\partial_{1\mu}\partial_{1\nu} + \partial_{2\mu}\partial_{2\nu}\Big)\nonumber\\
&\phantom{=}+ g_{\mu\nu}\Big(\square_1  + \square_2 \Big)-3g_{\mu\nu}m^2-g_{\mu\nu}\partial_{1\alpha}\partial_2^\alpha \Big]G_{13}G_{23}\, ,
\end{align}
which is 
\begin{align}
&\braket{T_{\mu\nu}(x_3)\Theta(x_1)\Theta(x_2)}=\frac{4m^4}{3}\int \frac{d^4p_1}{(2\pi)^4}\frac{d^4p_2}{(2\pi)^4}\frac{d^4p_3}{(2\pi)^4}\frac{e^{ip_1x_{12}+ip_2x_{13}+ip_3x_{23}}}{(p_1^2+m^2)(p_2^2+m^2)(p_3^2+m^2)}\\
&\quad \Big[-4p_{2(\mu}p_{3\nu)}+(p_{2\mu}p_{2\nu}+ p_{3\mu}p_{3\nu})-g_{\mu\nu}(p_2^2+p_3^2)-3g_{\mu\nu}m^2+ g_{\mu\nu}p_2\cdot p_3\Big]\, .\nonumber
\end{align}
The Fourier transform is 
\begin{align}
&\lll T_{\mu\nu}(\K_3)\Theta(\K_2)\Theta(\K_1)\rr =-g_{\mu\nu} m^4\int \frac{d^4p}{(2\pi)^4}\frac{\frac{4}{3}((p+\K_1)^2+(p-\K_2)^2)+4m^2+ \frac{4}{3}(p+\K_1) (p-\K_2)}{(p^2+m^2)((p+\K_1)^2+m^2)((p-\K_2)^2+m^2)}\nonumber\\
&\quad + \frac{4}{3}m^4\int\frac{d^4p}{(2\pi)^4} \frac{\Big[(p_{(\mu}+\K_{1(\mu})(p_{\nu)}-\K_{2\nu)})+(p_{\mu}+\K_{1\mu})(p_{\nu}+\K_{1\nu})+ (p_\mu-\K_{1\mu})(p_\nu-\K_{1\nu})
\Big]}{(p^2+m^2)((p+\K_1)^2+m^2)((p-\K_2)^2+m^2)}\, .\nonumber
\end{align}
The Fourier transform is 
\begin{align}
&\lll T_{\mu\nu}(\K_3)\Theta(\K_2)\Theta(\K_1)\rr =-\frac{4g_{\mu\nu} m^4}{3}\int \frac{d^4p}{(2\pi)^4}\frac{((p+\K_1)^2+(p-\K_2)^2)+m^2+ \frac{4}{3}(p+\K_1) (p-\K_2)}{(p^2+m^2)((p+\K_1)^2+m^2)((p-\K_2)^2+m^2)}\nonumber\\
&\quad + \frac{4}{3}m^4\int\frac{d^4p}{(2\pi)^4} \frac{\Big[(p_{(\mu}+\K_{1(\mu})(p_{\nu)}-\K_{2\nu)})+(p_{\mu}+\K_{1\mu})(p_{\nu}+\K_{1\nu})+ (p_\mu-\K_{1\mu})(p_\nu-\K_{1\nu})
\Big]}{(p^2+m^2)((p+\K_1)^2+m^2)((p-\K_2)^2+m^2)}\, .\nonumber
\end{align}
Evaluating this loop integral, we obtain
\begin{align}
\left.\lll T_{\mu\nu}(\K_3)\Theta(\K_2)\Theta(\K_1)\rr\right|_{O(\K^4)}&=
 \frac{\K_1^2}{720\pi^2}\left(\K_{1\mu}\K_{1\nu}+5\K_{1(\mu}\K_{2\nu)}\right)+ \frac{\K_2^2}{720\pi^2}\left(\K_{2\mu}\K_{2\nu}+5\K_{1(\mu}\K_{2\nu)}\right)\nonumber\\
&+\frac{\K_1\cdot \K_2}{720\pi^2}\left(\K_{1\mu}\K_{1\nu}+ \K_{2\mu}\K_{2\nu}+4\K_{1(\mu}K_{2\nu)}\right)\\
& -\frac{g_{\mu\nu}}{288\pi^2}\left(\K_1^4+\K_2^4+\K_1^2\K_2^2+\frac{9}{5} (\K_1^2+\K_2^2)\K_1\cdot\K_2 +\frac{6}{5}(\K_1\cdot\K_2)^2\right) \, ,\nonumber
\end{align}
where we only wrote the terms up to $O(\K^4)$. Note that specializing to $\mu=\nu=u$, we obtain the three-point function \eqref{tuuscalar} written in a more compact form:
\begin{align}
\left.\lll T_{uu}(\K_3)\Theta(\K_2)\Theta(\K_1)\rr\right|_{O(\K^4)}&=
 \frac{\K_1^2}{720\pi^2}\left(\K_{1u}^2+5\K_{1u}\K_{2u}\right)+ \frac{\K_2^2}{720\pi^2}\left(\K_{2u}^2+5\K_{1u}\K_{2u}\right)\nonumber\\
&+\frac{\K_1\cdot \K_2}{720\pi^2}\left(\K_{1u}^2+ \K_{2u}^2+4\K_{1u}K_{2u}\right)\, .
\end{align}

\addcontentsline{toc}{section}{References}
\bibliographystyle{utphys}
{\small
\bibliography{ref}
}

\end{document}